\documentclass[11pt]{article}
\pdfoutput=1

\usepackage{jheppub}
\usepackage[T1]{fontenc}
\usepackage{amssymb,amsthm}
\usepackage{tikz}
\usepackage{multirow}
\usepackage{algorithm,algpseudocode}
\usepackage{afterpage}

\usetikzlibrary{decorations.markings}
\tikzset{->-/.style={
    decoration={
        markings,
        mark=at position 0.5 with {\arrow{>}}
    },
    postaction={decorate}
}}

\newcommand{\Z}{\mathbb{Z}}

\title{Snowmass White Paper: \\ String Theory and Particle Physics}

\author[a,b,c]{Mirjam Cveti{\v c},}
\author[d,e]{James Halverson,}
\author[f]{Gary Shiu,}
\author[g]{and Washington Taylor}

\affiliation[a]{Department of Physics and Astronomy, and Department of Mathematics,\\ University of Pennsylvania,  
				Philadelphia, PA 19104-6396, USA}

	\affiliation[b]{Center for Applied Mathematics and Theoretical Physics, University of Maribor,
				Maribor, Slovenia}
				\affiliation[c]{Theory Department, CERN, CH-1211, Switzerland}
\affiliation[d]{Department of Physics, Northeastern University, Boston, MA 02115}
\affiliation[e]{The NSF AI Institute for Artificial Intelligence and Fundamental Interactions}
\affiliation[f]{Department of Physics, University of Wisconsin-Madison,\\ 1150 University Ave, Madison WI 53706, U.S.A.}
\affiliation[g]{Center for Theoretical Physics,
Department of Physics,
Massachusetts Institute of Technology\\
77 Massachusetts Avenue,
Cambridge, MA 02139, USA}

\emailAdd{cvetic at physics.upenn.edu,
j.halverson at northeastern.edu,  shiu at physics.wisc.edu, wati at mit.edu}

\abstract{
We review recent developments and outstanding questions  regarding
connecting the
top-down UV complete physical framework of string theory with
 the observed physics of the Standard Model and beyond the standard
 model physics, emphasizing the global nonperturbative framework of
 F-theory and general lessons from UV physics.
This paper, prepared for the TF01 conveners of the Snowmass 2022 process, provides a brief synopsis of this important area, focusing on ongoing developments and opportunities.  
}

\begin{document} 
\vspace*{-1cm}
	\begin{flushright}
	{\texttt{CERN-TH-2022-054}}\\
		{\texttt{UPR-1318-T}}\\
		{\texttt{MIT-CTP-5419}}\\
	\end{flushright}
\maketitle
\flushbottom

\section{Introduction}

Ever since its formative years, string theory has been developed in
close tandem with particle physics. Originally proposed as a theory of
the strong interaction, string theory owes its roots to particle
physics.  While string theory was soon after elevated to a quantum
theory of gravity thanks to the ubiquitous massless spin two particle
that the theory predicts, its connection to particle physics has only
grown stronger via the construction of increasingly realistic models and an understanding of typical features. The quest for unification of fundamental forces calls
for an ultraviolet-complete theory that simultaneously incorporates
both quantum gravity and chiral gauge theories.  The need to develop a
consistent framework to unify particle physics with gravity was the
backdrop for the heterotic string \cite{Gross:1984dd} and Calabi-Yau
compactifications \cite{Candelas:1985en}. The connection between
string theory and particle physics has evolved a great deal in the
past few decades. This invited white paper provides a synopsis of this
important area, often referred to as ``string phenomenology'',
focusing on ongoing developments and opportunities.

Some of the central questions in this area of research are:

\noindent
{\bf 1) How precisely can we match observable particle physics with
  specific string vacuum constructions?}
This question has motivated constructions of string solutions for
several decades, with a variety of different approaches leading to
string models that capture the gauge group, matter content, and
various other features of the observed Standard Model to varying
degrees of precision, usually in the context of supersymmetric
extensions.  In recent years more systematic approaches have given a
more global picture of the set of possibilities and enormous classes
of candidate (supersymmetric) Standard Model-like constructions,
though there are still many challenges in constructing
 a set of vacua that contain all observed features of the
Standard Model including Yukawa couplings, the Higgs field,
supersymmetry breaking, the
detailed structure of Standard Model parameters, and the observed cosmological constant.

\noindent
{\bf 2) What can be ruled out from string theory?}  While for some
time there was an outlook 
%a feeling 
in the community that ``anything goes'' in
string theory, it is clear that UV consistency with quantum gravity
imposes fairly stringent constraints on the set of allowed models of
quantum field theory available from string theory.  These constraints
are clearest  with more dimensions and supersymmetry, where for example
in 6D and 
10D
the size and complexity of the gauge group and matter
representations are already 
constrained by gravitational
anomalies.  These anomaly constraints can also be understood as
manifestations of geometric constraints in the UV; similar constraints
also hold in
% 4D
 other dimensions although they are less well understood
in 4D, and the study
of other constraints from UV physics is a very active area of current
research.

\noindent
{\bf 3) What features of string vacua are typical, and what are the
  consequences for particle physics?} While the set of possible consistent string
vacuum solutions is enormous, it is believed to be effectively finite.
And within this large finite set some features occur overwhelmingly
more often than others, at the level of naive counting of discrete sets (although a precise formulation of the measure across string vacua is still lacking).  As our understanding of the global set of
solutions of the theory increases we can make more informed
observations of what kinds of features, beyond the Standard Model, such as hidden gauge and matter
sectors, axion fields, etc., are typical in ``most'' string vacua.
Other features, while realized in some string vacua, may be highly
atypical and involve extensive fine-tuning, in a way that can be
quantified in terms of numbers of continuous moduli or discrete
parameters like fluxes that must take special values to realize these
features.  A crucial question for the next decade(s) is to identify
more precisely what kinds of additional beyond the Standard Model
structure are associated with typical realizations of Standard Model
and Standard Model-like structure in string vacua, and what features
are atypical or involve fine tuning. 

Beyond these guiding questions, string theory has also more generally
been a constant source of ideas for particle phenomenology. One of the
deepest puzzles facing particle physics today is the electroweak
hierarchy problem. Understanding the huge disparity between the weak
scale and the Planck scale has been the driving force for particle
theory in the past few decades. Given the ultraviolet sensitivity of
this question, a theory valid all the way to the Planck scale is
likely needed to make significant progress. As we review below, string
theory has shown to be resourceful not only in realizing existing
scenarios of physics beyond the Standard Model (such as supersymmetry,
axions, etc.) but suggesting new ones, as well as offering new insights
to the notion of naturalness. Only with a consistent quantum theory of
gravity can we address other vexing naturalness problem involving
gravity, such as the unreasonable smallness of the cosmological
constant.
 
The subjects of particle physics and cosmology are deeply
intertwined. We focus on particle physics aspects in this
contribution, leaving the connection between string theory and
cosmology to a complementary white paper \cite{Flauger:2022hie}.

We should emphasize that the emphasis on topics in this contribution
 is influenced by
the expertise and research interests of the authors; while we have
made an effort to summarize the state of the field in the areas
connecting string theory and particle physics that we believe are of
most current significance and have greatest potential for developments
in the coming years, we have not attempted to be comprehensive and
some recent related developments are treated briefly or not at all.

%
%[Complementary to \cite{Snowmass_Cosmology}. Emphasis on particle
%  physics aspects though particle physics and cosmology are deeply
%  connected.]
%
%
%[What has been accomplished, what are the ongoing developments, and
%  what remains to be done.]

\section{Particle Physics from String Theory}

Finding a realization of string theory solutions with the structure of
the Standard Model gauge group and three generations of Standard Model chiral matter
fields, as well as additional features such as the Higgs sector and
proper Yukawa
couplings,  moduli stabilization, supersymmetry breaking
and 
%vanishing 
vanishingly small positive cosmological constant, etc.
has been a major
preoccupation of string theorists over the last three decades. 
For many years such constructions were developed on a somewhat ad hoc
basis using a variety of available tools, primarily in the perturbative corners of string theory, such as algebraic geometry techniques in heterotic string theory and  orbifold
constructions in Type II string  theory, and Standard Model-like constructions seemed
somewhat rare.  However, developments in recent years have given a better global
picture of large parts of the 
%space
landscape 
of string vacua.  There is
increasing evidence that there are huge classes of string vacua with
Standard Model-like features, and, at least for vacuum
solutions with supersymmetry there are increasingly powerful
approaches to systematically identifying how such Standard Model-like
vacua fit into the landscape, and how natural such solutions are. In
particular, the approach of F-theory \cite{VafaF-theory,MorrisonVafaI,MorrisonVafaII} provides a geometric
framework in which the broadest class of supersymmetric string vacua
yet identified are incorporated in an underlying connected moduli space
(including most heterotic vacua as duals of a subset of F-theory vacua).
While F-theory is intrinsically nonperturbative, the global lessons
from this approach provide insight into the structure of the overall
landscape and provide guidance for further work in narrowing down the
nature of Standard Model-like solutions, computing more detailed
features of these solutions, and identifying what physics beyond the
Standard Model may naturally arise in tandem with it.  The biggest
challenges at this point to this approach are understanding
supersymmetry breaking 
 (\S\ref{sec:SUSY-breaking})
and the related questions of moduli stabilization and the small
cosmological constant (\S\ref{sec:moduli}); these latter issues are addressed more
comprehensively in \cite{Flauger:2022hie}.

\subsection{Approaches to Standard Model-like constructions
  in string theory}
	
In the past few decades, enormous efforts have been undertaken to
demonstrate explicit top-down string theory constructions of vacua
with the gauge symmetry and particle spectrum of the Standard Model.
This program originated in the studies of the $E_8 \times E_8$
heterotic string compactified on Calabi-Yau threefolds
\cite{Candelas:1985en,Greene:1986ar,Braun:2005ux,Bouchard:2005ag,Bouchard:2006dn,Anderson:2009mh,Anderson:2011ns,Anderson:2012yf}.
Other efforts have involved Standard-like Model constructions on
heterotic orbifolds \cite{Lebedev:2007hv,Lebedev:2006kn} and free
fermionic constructions \cite{Faraggi:1989ka,Faraggi:1991jr}. The
first globally consistent construction with the exact matter spectrum
of the Minimal Supersymmetric Standard Model (MSSM) heterotic was
given in \cite{Bouchard:2005ag,Bouchard:2006dn}, and more recently
sophisticated computational and mathematical analyses of line bundle
constructions have led to the construction of thousands of vacua with
the Standard Model gauge group and three generations of Standard Model chiral matter
fields based on the heterotic approach \cite{Anderson:2011ns,
  Anderson:2013xka}.  With the advent of D-branes
\cite{Polchinski:1995mt} these efforts were further advanced by
studying D-branes at singularities (see
\cite{Shiu:1998pa,Aldazabal:2000sa,Cvetic:2000st, Berenstein:2001nk, Verlinde:2005jr} and references
therein) and intersecting D-brane models in type II string theory
\cite{Berkooz:1996km,Aldazabal:2000dg,Aldazabal:2000cn,Ibanez:2001nd,Blumenhagen:2001te,Cvetic:2001tj,Cvetic:2001nr}
(for review see \cite{Blumenhagen:2005mu} and references therein),
which led to the first globally consistent three-family Standard-like
Model constructions \cite{Cvetic:2001tj,Cvetic:2001nr}.  Both
Heterotic and Type II constructions produced large classes of globally
consistent models with Standard Model gauge sectors and three chiral
families. One should, however, point out, that these constructions may
be limited, in part, due to the perturbative values of the string
coupling in these approaches.  Furthermore, these models typically
suffer from chiral and vector-like exotic matter, though by now there
are large classes of constructions that are increasingly close to the MSSM.
Another approach to string compactification is based on using special
holonomy $G_2$ seven-manifolds to compactified M-theory to 4
dimensions.  There has been substantial progress in this direction in
recent years, as discussed further in \S\ref{sec:math}; although a
full description of a model with the Standard Model gauge group and
chiral matter content in terms of a singular $G_2$ geometry is still
some way in the future (for some initial developments in this
direction, see
 \cite{Acharya:2008zi}), this may in time be a promising approach for
construction of a large class of semi-realistic string vacuum models.

While the above mentioned constructions concentrated on perturbative
corners of string theory, 
the geometric approach of F-theory \cite{Vafa:1996xn,oai:arXiv.org:hep-th/9602114,oai:arXiv.org:hep-th/9603161} 
gives a systematic and nonperturbative
global picture of an even larger class of nonperturbative string
vacua.  
%F-theory
%\cite{Vafa:1996xn,oai:arXiv.org:hep-th/9602114,oai:arXiv.org:hep-th/9603161}.
In F-theory, the
back-reactions of non-perturbative 7-branes onto the geometry of
six compactified space dimensions are encoded in the geometry
of an elliptically fibered Calabi-Yau fourfold. 
By studying this
space of theories with well-established tools of algebraic geometry, 
a variety of
top-down F-theory constructions have been realized where the gauge
degrees
of freedom are encoded in
the singularity structure of the elliptically fibered Calabi-Yau
fourfold. 
Following the initial intensive study of GUT F-theory models
based on tuned SU(5) GUT groups initiated by
\cite{DonagiWijnholtGUTs,BeasleyHeckmanVafaI,BeasleyHeckmanVafaII,DonagiWijnholtModelBuilding},
a number of other classes of F-theory  constructions
with the Standard Model gauge group
have been realized including directly realizing the Standard Model gauge group as
a geometrically rigid symmetry \cite{GrassiHalversonShanesonTaylor},
geometrically tuned Standard Model gauge groups \cite{Klevers:2014bqa,Cvetic:2015txa,
Taylor:2019wnm,Raghuram:2019efb}, and Standard Model constructions from
flux breaking of exceptional GUTs \cite{Li:2021eyn}.  

In F-theory, the matter spectrum is uniquely fixed by a background
gauge  configuration, which can be conveniently specified by the three-form
gauge potential $C_3$ in the dual M-theory geometry.  The chiral
spectrum depends only on the field strength $G_4 = dC_3$, referred to
as flux.  By now, there exists an extensive toolbox for constructing
and enumerating the $G_4$ flux configurations
\cite{Marsano:2010ix,Braun:2011zm,Marsano:2011hv,oai:arXiv.org:1111.1232,Krause:2012yh,Braun:2013nqa,Cvetic:2013uta,Cvetic:2015txa,Lin:2015qsa,Lin:2016vus,Jefferson:2021bid}.
The application of these tools has led to the construction of a
variety of globally consistent chiral F-theory particle physics
constructions
\cite{Krause:2011xj,Cvetic:2015txa,Lin:2016vus,Cvetic:2018ryq},
particularly in the class of tuned Standard Model gauge groups, which recently
culminated in the largest explicit class of string vacua that realize
the Standard Model gauge group along with the exact chiral spectrum
and gauge coupling unification \cite{Cvetic:2019gnh}.  
The construction of global Standard Model vacua from the  SU(5) GUT
approach is complicated by the necessity for including hypercharge
flux, which requires more complicated (non-toric) base manifolds
\cite{Mayrhofer:2013ara,Braun:2014pva}.
For a review of
earlier efforts on constructions of (non-compact) SU(5) GUTs, see reviews
\cite{Heckman:2010bq,Schafer-Nameki:2015bva}, and for a comprehensive introduction to F-theory compactification \cite{Weigand:2018rez}. 

The constructions just described begin to address the first question
raised in the Introduction: How precisely can we match observable
particle physics with specific string vacuum constructions?  In the
following subsections we address how much further these constructions
can go beyond the gauge group and chiral matter content of the Standard Model.
Related to the third question of the Introduction, however, there is
also a question of typicality.  As discussed further in
\S\ref{sec:ml}, the number of complex threefold base geometries that
support elliptically fibered Calabi-Yau fourfolds is enormous, likely on the order
of $10^{3000}$.  One current area of active research is to understand
which of the above constructions are most typical (involve the least
fine tuning).  While the (tuned) SU(5) GUT models have been studied in
the most detail,  these models seem to be possible only on a small subset of
the allowed F-theory bases and involve extensive tuning of moduli
\cite{BraunWatariGenerations}.  Similar issues hold for the 
constructions in which the Standard Model gauge group is directly tuned at special
loci in the generic Weierstrass model over the base, as in the
approach taken in
\cite{Klevers:2014bqa,Cvetic:2015txa,Raghuram:2019efb,Cvetic:2019gnh}.
In terms of simple numerical counting, all but a handful of the
enormous number of threefold bases that support F-theory constructions
are populated by numerous geometric gauge factors associated with
rigid divisors in the base; in 4D, these rigid gauge groups include
the non-Abelian factors
 $E_8, E_7, E_6, F_4, G_2, SO(8), SO(7), SU(3), SU(2)$
but not e.g.\ $SU(5)$ or $SO(10)$ \cite{MorrisonTaylor4DClusters}.  It is natural to expect that the
most typical realizations of the MSSM in F-theory will arise through
these rigid gauge groups, as explored in
e.g. \cite{TianWangEString,Li:2021eyn}.  While there are many
challenges associated with making precise sense of the measure problem
on the space of flux compactifications (see
e.g.\ \cite{Denef:2004ze,DenefLesHouches}), the enormous numbers of
vacua involved in geometric F-theory constructions suggest  that some
insight can be gained on questions of typicality of standard model
constructions; further work in coming years should address how the
more detailed physics of the different constructions described here
differs and what physics beyond the Standard Model is more or less typical.

\subsection{Matter fields and the Standard Model}

As described above, we now have the technical facility to construct
enormous classes of models with the Standard Model gauge group and
chiral matter content.  The next challenges are to understand more
clearly the more detailed structure of these models.  
While F-theory provides a powerful nonperturbative framework for
accessing the large scale picture of the set of vacua, explicit
calculations of more detailed aspects of these vacua require technical
tools not yet developed and which in some cases  may be hard to access
due to the fundamental nonperturbative nature of the physics in these
solutions.  Current work is focused on understanding more detailed
aspects of the vector-like part of the matter content, Higgs fields
and Yukawa interactions.  

In particular, the methods described above
are insufficient to determine the exact vector-like spectrum of the
chiral zero modes, which depend not only on the flux $G_4$, but also
on the flat directions of the three-form potential $C_3$. In
\cite{Bies:2014sra,Bies:2017fam}, methods for determining
the exact vector-like spectra were put forward, and further advanced in the context of Standard Models in \cite{Bies:2021nje,Bies:2021nje}.  Attaining a more
detailed understanding of the vector-like spectrum and Higgs fields
is a key goal of
near-term research in this area.  

Yukawa couplings in F-theory have been explicitly computed only in the ultra-local F-theory
models \cite{Heckman:2008qa,Hayashi:2009ge,Cecotti:2009zf,Marsano:2011hv}.
 However, there has been recent progress on calculations of the
 holomorphic part of Yukawa couplings within a global SU(5) GUT-like
 model \cite{Cvetic:2019sgs}.  
The connection between F-theory and type II descriptions of Yukawa
couplings has also been elucidated \cite{Collinucci:2016hgh} via the
role of D-instanon contributions \cite{Blumenhagen:2006xt,Ibanez:2006da,Florea:2006si,Blumenhagen:2007zk}.(For review, see \cite{Blumenhagen:2009qh}.)
Furthermore the calculation of the
 K\"ahler potential, which determines the normalization of the kinetic
 energy terms and thus needed for the physical values of Yukawa
 couplings, remains an outstanding problem. Note that on the latter
 topic progress has been made in the heterotic
 \cite{Kakushadze:1997ub,Braun:2008jp} and Type II context
 \cite{Cvetic:2003ch, Cvetic:2009yh}. The full determination of
 physical Yukawa couplings in globally consistent F-theory Standard
 Model constructions is an important goal of future research, and
 affects aspects such as the possibility of proton decay in these
 constructions.

Another class of questions relates to how typical the matter content
of the Standard Model is among string vacua.  While the light chiral
matter content of the Standard Model, including 3 generations of
quarks and leptons, appears somewhat arbitrary, and many string
constructions that give the Standard Model gauge group also include
various exotic matter fields, there are also strong constraints from
anomaly cancellation conditions that limit the possible sets of matter
fields.  Naively, however, the matter that occurs in nature could live in
arbitrarily complicated high-dimensional/high-charge representations
of the non-Abelian and Abelian factors in the Standard Model gauge group $SU(3)
\times SU(2) \times U(1)$ while still satisfying anomaly cancellation.
There are some indications, however, that
the set of possible matter representations realized in string theory
is quite limited, so that  only special simple
representations like those of the Standard Model are realizable in a
UV complete theory.
The range of allowed representations for heterotic orbifold models was studied
some time ago in \cite{Dienes:1996yh}.  More recently, analysis from the
F-theory point of view seems to strongly limit the range of possible
matter representations of non-Abelian groups that may arise from
geometry \cite{Klevers:2017aku,Cvetic:2018xaq,Esole:2020tby}, although there is more apparent flexibility for U(1)
charges \cite{Taylor:2018khc,Raghuram:2018hjn,Cianci:2018vwv,Collinucci:2019fnh,Raghuram:2021wvx}.
Furthermore, the global structure of the gauge group affects the set
of allowed matter fields; the observed Standard Model chiral matter
spectrum is much more natural if the global structure of the group is
$(SU(3) \times SU(2) \times U(1))/\Z_6$ (as occurs in most GUT models
and  many of the F-theory constructions delineated above) than the
group without the quotient;  in the F-theory context these issues have been studied recently in
\cite{Cvetic:2017epq,Taylor:2019ots,Taylor:2019wnm,Raghuram:2019efb}\footnote{In F-theory the global structure of a gauge group is determined by the torsional part of the Mordell-Weil group of an elliptic fibration, addressed in non-Abelian cases  in \cite{Mayrhofer:2014opa} and in the presence of Abelian factors in  \cite{Cvetic:2017epq}. For review, see \cite{Cvetic:2018bni}.}.  Indeed, the ``completeness
hypothesis'' (see e.g.\ \cite{Banks:2010zn}), recently
proven in AdS space in \cite{Harlow:2018tng}, would indicate that if the gauge group
lacks the quotient there must be massive exotics with nontrivial
charge under the central $\Z_6$.
If string theory can be shown to strongly limit the set of
possible matter fields that can appear in a low-energy vacuum with the
gauge group of the Standard Model, it will help to fit observed
physics into the landscape of ``typical'' theories, while otherwise
there is another fine-tuning problem.  Related issues are also
discussed in \S\ref{sec:UV}.

Beyond the light chiral matter fields in the Standard Model, many string constructions
contain a variety of further matter fields that may play roles as dark
matter candidates; we explore these further in the following subsection.

\subsection{Particle Remnants of the String Landscape}

As mentioned above,
string compactifications that give rise to the gauge group
and chiral matter content of the Standard Model regularly exhibit new
particles that are accidental consequences of the ultraviolet theory,
i.e., they are not motivated by shortcomings of the Standard Models of
particle physics or cosmology. Nevertheless, these particle remnants
are potentially observable, providing both experimental constraints
and opportunities. See \cite{Halverson:2018vbo} for a thorough account
in lectures, including the topics discussed below: axions, dark gauge
sectors, and vector-like exotics.

\vspace{.2cm}
\noindent \textbf{Axions.} String theory contains higher-form gauge
fields that regularly give rise to axions upon compactification
\cite{Svrcek:2006yi}; in supersymmetric models, these axions are  part of
the same multiplet as scalar moduli fields (\S\ref{sec:moduli}).
The
 number of axions $N$ is dictated by
the topology of the extra dimensions. Generally, $N$ is quite large;
in the largest ensembles of supersymmetric type IIB or F-theory
compactifications studied to date, $N$ is in the hundreds
\cite{Kreuzer:2000xy} or thousands
\cite{Taylor:2015xtz,Taylor:2015ppa,Halverson:2017ffz,Taylor:2017yqr},
respectively. This leads to the expectation that string theory gives
rise not to one or a few axions, but an axiverse
\cite{Arvanitaki:2009fg}, with diverse set of implications for
particle physics and cosmology; notably, these axions need not be the
QCD axion. Furthermore, recent results demonstrate \cite{Demirtas:2018akl} 
that type IIB compactifications with large volume and weak coupling typically exhibit 
numerous very light axions, well below the eV scale.
Possible implications of string theory for axions include
for solving the strong CP problem \cite{Demirtas:2018akl}, axion monodromy
inflation
\cite{McAllister:2008hb,Flauger:2009ab,Marchesano:2014mla,Hebecker:2014eua,McAllister:2014mpa,Hebecker:2014kva}
and reheating \cite{Blumenhagen:2014gta,Halverson:2019kna},
cosmological relaxation of the weak scale (aka the ``relaxion''
scenario) \cite{Ibanez:2015fcv,Brown:2016nqt,McAllister:2016vzi},
fuzzy dark matter \cite{Hui:2016ltb,
  Halverson:2017deq,Cicoli:2021gss}, black hole superradiance
\cite{Mehta:2021pwf}, gravitational waves
\cite{Kitajima:2018zco,Barnaby:2012xt,Mukohyama:2014gba,Liu:2017hua,Liu:2018rrt}, and
couplings to axion-like particles \cite{Halverson:2019cmy,
  Demirtas:2021gsq}. 

\vspace{.2cm}
\noindent \textbf{Dark Gauge Sectors.} Consistency conditions such as
Gauss' law for higher-form gauge fields, generally associated with
tadpole cancellation conditions, combined with geometric and
topological features in the UV theory often require the presence of
multiple gauge sectors, which may couple to the visible sector (if present)
only via gravity and non-renormalizable interactions.
For instance,
visible sectors arise in one $E_8$ factor in the heterotic string,
leaving the other $E_8$ as a potential dark matter sector
\cite{Faraggi:2000pv}, where some gauge factor may be forced from a
given distribution of instantons. More generally, as discussed above, F-theory
compactifications typically give rise to dozens or
hundreds of rigid gauge
factors. 
These heterotic and F-theory mechanisms producing hidden sectors are
related through duality \cite{Friedman:1997yq}; duality with $G_2$ constructions
suggests similar features there \cite{Braun:2017uku}.
In many cases these extra sectors
confine, giving rise to the possibility
of dark glueball dark matter
\cite{Halverson:2016nfq,Dienes:2016vei,Acharya:2017szw}. However, 
%in that
 it is very easy to oversaturate the observed relic abundance
\cite{Halverson:2016nfq}, and more generally one might use
cosmological observations to constrain F-theory, given the plethora of
dark gauge sectors and axions that it exhibits. In addition, string
theory also suggests new portals between the visible and dark gauge
sectors e.g., through Stuckelberg couplings \cite{Feng:2014cla,
  Feng:2014eja}. 
  
\vspace{.2cm}
\noindent \textbf{Vector-like Exotics.}  Matter fields that are
vector-like with respect to the Standard Model regularly arise in
string compactifications and are experimentally allowed, provided that
their mass is TeV-scale or above. Such new particles are the subject
of many searches at LHC, and provide
opportunities to explain existing data, such as WIMP dark matter
\cite{Cirelli:2005uq,Halverson:2014nwa} and the B anomalies
\cite{LHCb:2021trn}. In string constructions a variant of vector-like
exotics regularly arise, so-called quasi-chiral exotics
\cite{Anastasopoulos:2006da,Cvetic:2011iq,Halverson:2013ska}, which
are vector-like with respect to the Standard Model, but chiral with
respect to another symmetry, which provides a mechanism to expect
masses significantly below the string scale. In some other cases,
these quasi-chiral exotics decouple from the low energy spectrum due
to hidden sector strong dynamics and charge confinement
\cite{Kakushadze:1997mc,Cvetic:2002qa}.

\medskip

\subsection{SUSY breaking}
\label{sec:SUSY-breaking}

One of the greatest challenges in string phenomenology is the issue
that the vacua that are best understood are those with at least some
supersymmetry.  Since low-scale supersymmetry (i.e. SUSY at the TeV
scale or below) has not been observed at the LHC (or in astrophysical
experiments), this presents a major challenge to connecting UV
complete string constructions with observable physics.  The most
powerful approaches to understanding string vacua and computing
physics in these backgrounds relies heavily on supersymmetry, in
particular the nonperturbative approach of F-theory uses algebraic
geometric methods in which the complex-analytic structure of the
geometry is tied to space-time supersymmetry.  Certainly, direct
exploration of non-SUSY vacua and mechanisms of high-scale SUSY
breaking in supersymmetric string vacua are significant priorities for
this community.
The problem of identifying non-supersymmetric de Sitter
string vacua from conventional methods such as uplifted flux vacua is
a major focus of current research, discussed slightly further in the
following section and in more depth in \cite{Flauger:2022hie}.

At the same time, some lessons even from supersymmetric string vacua
such as the presence of many scalar fields and axions, and
gravitationally coupled hidden dark sectors from rigid gauge groups
and associated matter uncharged under the Standard Model gauge group,
seem to be present in the string landscape somewhat independently of
the amount of supersymmetry involved. For example, virtually the same
hidden sector rigid gauge group factors (e.g. $E_8, E_7, E_6, F_4,
SO(8), G_2, SU(3), SU(2)$ but not $SU(5)$) arise in 6D supersymmetric
F-theory vacua with 8 supercharges as in 4D supersymmetric F-theory
vacua with 4 supercharges
\cite{MorrisonTaylorClusters,MorrisonTaylor4DClusters}, and with
similar frequencies
\cite{Taylor:2015ppa,Halverson:2017ffz,Taylor:2017yqr}, suggesting
that some of these qualitative features of string vacua may persist
even without SUSY or when SUSY is broken.

Another intriguing question is whether the geometries associated with
supersymmetry are natural even if we drop the constraint of SUSY.  In
particular, there are few smooth compact manifolds known in the math
literature that admit a Ricci flat global metric that are not
essentially Calabi-Yau manifolds or closely related special holonomy
manifolds \cite{Acharya:2019mcu}.  (As a very simple example, in 2D
the Gauss-Bonnet theorem shows that the only compact smooth 2D
manifold admitting a Ricci flat metric is the torus.)  Indeed, it was
conjectured in \cite{Acharya:2019mcu} that any smooth manifold with a
Ricci flat metric that lacks special holonomy leads to a ``bubble of
nothing'' instability \cite{Witten:1981gj}, and there is some recent
evidence in support of this
\cite{GarciaEtxebarria:2020xsr,Acharya:2020hsc}.  
%This is a special
%case of the more general hypothesis, discussed further in the
%following section, that de Sitter vacua in general may be unstable.
This is a special case of the more general hypothesis that
non-supersymmetric vacua may suffer from some form of instability. We
will return to this in the next section when we discuss de Sitter
vacua.  Another approach to realizing non-supersymmetric vacua is to
start with a hyperbolic compactification space and directly construct
a solution with positive cosmological constant, as explored for
example in \cite{DeLuca:2021pej} and references therein.  The effort
to directly construct explicit non-supersymmetric string vacua is a
promising avenue for further research that might justify (or might
disprove) the notion that features of supersymmetric string vacua are
also natural for our non-supersymmetric world.

\subsection{Moduli Stabilization}
\label{sec:moduli}

String theory has no free parameters;
for example in type II string theory,
 the string coupling is encoded in the dynamical
axiodilaton field.
 Upon compactification, the low-energy effective action of 
string theory constructions generically contains many
parameters associated with vacuum expectation values (VEVs) of scalar
fields, generalizing  the type II axiodilaton field.
Stabilizing these scalar fields (known as moduli) is thus important for drawing phenomenological predictions from string theory.
Because no massless scalar fields are observed in nature,
we expect that all these moduli are stabilized in a physically
relevant vacuum of the theory.  Fifth-force experiments and cosmological constraints typically put these moduli masses above ${\cal O} (10)$ TeV scale, though loopholes exist if one contemplates unconventional cosmological histories.
Understanding precisely how moduli stabilization
occurs is a major challenge for string theory. 
In some cases, the issue of moduli stabilization cannot be decoupled from particle physics considerations \cite{Blumenhagen:2007sm}, though model-building can mitigate this tension \cite{Blumenhagen:2009gk,Grimm:2011tb,Cicoli:2012vw}.
In general, it is expected that any theory with broken supersymmetry
will generically have mass terms for all the scalar fields. The question remains why the Higgs mass is significantly lighter than that of other moduli. This issue is akin to realizing inflation in string theory, which requires a hierarchy in masses between the inflaton and other moduli.

Mechanisms to stabilize moduli developed so far can be divided into
two broad classes: power-law stabilization and non-perturbative
stabilization. We refer the reader to \cite{Flauger:2022hie} for their
distinction and a more in-depth discussion. Here, we focus on the
latter as they are more commonly used in particle physics model
building from string theory.  In general, even constructions of
supersymmetric string vacua incorporate mechanisms that stabilize most
or all moduli fields.  In ${\cal N} = 1$ solutions of type II string
theory and F-theory, for example, fluxes generate a superpotential
that stabilizes many of the (complex structure) moduli of the vacua
directly (although there are questions about whether the number of
fluxes, which is bounded by the D3-brane tadpole, is sufficient to
stabilize a plurality of the complex structure 
moduli 
\cite{Carta:2019rhx,Bena:2020xrh,Bena:2021wyr,Gao:2022fdi}); other effects such as gaugino condensation and
nonperturbative instanton affects also arise that in general are
expected to lift other moduli of the theory.  See, e.g.,
\cite{Douglas:2006es} for a review\footnote{Well-studied approaches to non-perturbative moduli stabilization include the KKLT scenario \cite{Kachru:2003aw} and the Large Volume Scenario \cite{Balasubramanian:2005zx}.}.  While this gives rise to a
plethora of expected vacua with negative cosmological constants
(including supersymmetric vacua), it is remarkably difficult to
explicitly construct moduli-stabilized vacua with positive
cosmological constant.  There are a number of no-go theorems that
demonstrate the impossibility of getting vacua with a positive
cosmological constant (or even a slowly rolling vacuum with many
e-folds of inflation) with simple (albeit limited) sets of ingredients
from supergravity theory
\cite{Maldacena:2000mw,Hertzberg:2007wc,Danielsson:2009ff,Wrase:2010ew}.
More involved setups that naively evade these no-go theorems were found to suffer from tachyonic instabilities \cite{Danielsson:2011au,Shiu:2011zt,Danielsson:2012et,Junghans:2016uvg,Junghans:2016abx} (an issue later revisited in e.g. \cite{Andriot:2018wzk,Andriot:2018ept,Garg:2018reu})).
It should be noted that the sources of potential considered in these no-go theorems
are not general enough to exclude de Sitter vacua in string theory.
It has been suggested that de Sitter vacua do not arise in regions of
string theory that are within strict parametric control
\cite{Danielsson:2018ztv,Obied:2018sgi,Ooguri:2018wrx}.  A lack of
parametric control does not mean that no theoretical control is
possible. However, in light of the Dine-Seiberg problem
\cite{Dine:1985he}, one would have to carefully quantify that higher
order terms which are ignored are indeed unimportant even though
different order terms in the potential compete to give a minimum.
In addition, the instability hypothesis  mentioned in
\S\ref{sec:SUSY-breaking} (see \cite{Freivogel:2016qwc} for a recent discussion
regarding de Sitter instability)
 suggests that any proposed de Sitter
vacuum should be subject to a more careful scrutiny.
Another perspective on flux vacua without supersymmetry is given in
\cite{Sethi:2017phn}.
  These various considerations have reignited
the interest in moduli stabilization in string theory, resulting in a
number of technical advances.  Recent work \cite{Demirtas:2021nlu} has
succeeded in constructing supersymmetric AdS vacua with exponentially
small negative cosmological constant, a difficult task even with an
optimized computer search \cite{Cole:2019enn}.  Challenges remain in
uplifting these solutions to de Sitter, all related one way or the
other to the anti-D3-branes that break supersymmetry. Metastability of
anti-D3-branes requires a large enough warped throat supported by a
large quantity of D3-brane charge which may be in further
tension with the
tadpole constraints mentioned above, though
detailed model building may mitigate this problem. It was argued that
a large throat leads to a singular bulk problem
\cite{Gao:2020xqh}. The singularities of the internal metric may be
resolved at the non-perturbative level \cite{Carta:2021lqg}, but
theoretical control of the low energy effective theory remains to be
shown. Furthermore, advances in lifting gaugino condensates to 10D
have recently been made, including an improved understanding of the
four-fermion couplings
\cite{Hamada:2018qef,Hamada:2019ack,Kachru:2019dvo,Hamada:2021ryq} and
the generalized complex geometry \cite{Kachru:2019dvo,Bena:2019mte}
needed to describe their backreaction to the internal geometry.  Such
10D lift allows us to quantify possible corrections to the stabilized
vacua which is a result of an interplay of the classical flux energy,
the non-perturbative instanton effects, and the anti-branes.  Fleshing
out these advances in full and constructing explicit vacua with all
moduli stabilized, particularly with a small positive cosmological
constant, remains a clear and precise challenge for the field in
coming years.

A complementary approach to explicitly stabilizing moduli is to study what is possible, in principle, in regimes of the theory that may be controlled. For instance, for increasing topological complexity, control over the $\alpha'$ and $g_s$ expansion in type IIB compactifications pushes the theory \cite{Demirtas:2018akl} to larger cycle volumes. This has numerous phenomenological implications; e.g., it correlates with increasingly light axions when their masses are generated non-perturbatively, and increasingly weak seven-brane gauge couplings. Applying such an analysis in the context of the $10^{15}$ F-theory compactifications with the gauge group and chiral spectrum of the Standard Model, it was found 
\cite{Cvetic:2020fkd} that (for the bulk of the models) there is no part of the large volume moduli space that realizes the correct values of the Standard Model gauge couplings; they are far too weakly coupled. Avoiding this consequence requires either alternative moduli stabilization schemes that exist at small volume, focusing on models with lower topological complexity, or modifying the location of the Standard Model sector in the extra dimensions, which forces the existence of dark gauge sectors. Results such as these are demonstrative of a general principle in string compactification: there is no free lunch, and correlations from the UV theory often exist that defy expectations from an EFT point of view.

\subsection{UV lessons}
\label{sec:UV}

The electroweak hierarchy problem is arguably one of the main science
drivers for particle physics beyond the Standard Model. In essence, it
is a question of how an infrared (IR) scale can emerge from an
ultraviolet (UV) scale without fine-tuning of UV parameters.  In
unifying particle physics with gravity, among the tunings in question
is the huge disparity between the Higgs mass and the Planck scale. A
notion of ``naturalness'' in accessing the degree of fine-tuning was
succinctly articulated by 't Hooft \cite{tHooft:1979rat}: ``at any
energy scale $\mu$, a physical parameter or set of parameters
$\alpha_i (\mu)$ is allowed to be very small only if the replacement
of $\alpha_i (\mu)=0$ would increase the symmetry of the system.''
There are reasons to expect that gravity may provide a break from this
Wilsonian EFT reasoning.  Heuristic arguments \cite{Banks:2010zn} as
well as string theoretical reasonings
\cite{Banks:1988yz,Harlow:2018tng} suggest that quantum theories of
gravity admit no global symmetries, making the above $\alpha_i (\mu)
\rightarrow 0$ limit subtle. Another hint is the Beckenstein-Hawking
entropy of a black hole, which links the classical IR solution of
gravity with the degeneracy of highly massive states of the UV
theory. This UV/IR mixing (a term coined in \cite{Minwalla:1999px})
manifests in many forms in string theory. There have been attempts
e.g. \cite{Abel:2021tyt} to frame the Higgs mass computation in string
theory in a way that respects this UV/IR duality. How this UV/IR
mixing can concretely address the electroweak hierarchy problem
remains to be explored. Meanwhile the vast but finite landscape of
string vacua also suggests a notion of stringy naturalness
\cite{Douglas:2004zg} which measures the degree of tuning by the
number of phenomenologically acceptable vacua leading to a given value
of an observable. Similar considerations of stringy naturalness may
shed light on the scale of supersymmetry breaking
\cite{Susskind:2004uv,Douglas:2004qg,Dine:2004is} as well as other
vexing hierarchy problems such as the smallness of the cosmological
constant, though concrete realizations remain to be found.

The Swampland program \cite{Vafa:2005ui} (see
e.g. \cite{Brennan:2017rbf,Palti:2019pca,vanBeest:2021lhn,Grana:2021zvf,Harlow:2022gzl}
for reviews) aims to make precise this UV/IR relation with an eye
toward its phenomenological implications. Combining reasoning from
such diverse areas as black hole physics, holography, scattering
amplitudes, and the bootstrap, an interconnected web of swampland
criteria has emerged. These criteria, if proven, may have interesting
phenomenological implications. For example, milli-charged dark matter
scenarios often considered in phenomenological studies are in tension
with the absence of global symmetry in quantum gravity
\cite{Shiu:2013wxa}. The Weak Gravity Conjecture (WGC)
\cite{Arkani-Hamed:2006emk} has been used to put phenomenological
constraints on axions \cite{Brown:2015iha, Brown:2015lia,
  Montero:2015ofa,Heidenreich:2015wga,Hebecker:2018ofv} and dark
photons \cite{Reece:2018zvv}. Stronger versions of the WGC have been
used to link the observed value of the cosmological constant with the
neutrino masses (which for fixed Yukawa couplings, set the weak scale)
\cite{Ibanez:2017kvh,Hamada:2017yji}. Ideas to stabilize the Higgs
mass using the WGC in the presence of scalars have been explored
\cite{Lust:2017wrl,Craig:2019fdy}. See the related Snowmass white
papers \cite{Draper:2022pvk, deRham:2022hpx} for discussions of other
phenomenological implications. While there is growing evidence for
various swampland criteria, they are at present conjectural though
continued serious attempts for proofs have been made
\cite{Cheung:2018cwt,Hamada:2018dde,Montero:2018fns,
  Bellazzini:2019xts,Arkani-Hamed:2021ajd}.

Swampland considerations have also been utilized to constrain the
gauge and matter content in consistent quantum theory of gravity. The
Completeness Hypothesis \cite{Polchinski:2003bq} necessitates physical
states with all possible gauge charges consistent with Dirac
quantization. In the presence of higher form symmetries, these
physical states include extended objects, such as strings and branes.
A powerful general approach initiated in \cite{Kim:2019vuc} makes use
of brane probes to rule out infinite families of anomaly free
gravitational theories.  Subsequent works have found precise match of
the allowed spectra with string constructions, sometimes involving
fine details such as the global structure of the spacetime gauge group
\cite{Cvetic:2020kuw,Montero:2020icj,Font:2020rsk,Cvetic:2021sjm,Font:2021uyw,Cvetic:2021sxm,Bedroya:2021fbu,Cvetic:2022uuu}.
Investigations along this line, if successfully extended to lower
dimensions, would suggest a notion of string universality
\cite{Adams:2010zy} (or the string lamppost principle
\cite{Montero:2020icj}) that all consistent supersymmetric theories of
quantum gravity are realized in string theory.  Six-dimensional
supergravity theories and their string/F-theory realizations provide
an excellent testbed for swampland ideas since the class of string
constructions is fairly well understood and controlled, and known
quantum consistency conditions also tightly bound the set of possible
low-energy theories; these theories satisfy the completeness relation
and some related conditions that may help to fully understand the role
of quantum gravity constraints in this context
\cite{Morrison:2021wuv,Raghuram:2020vxm}.  Improved understanding of
quantum gravity constraints in more general contexts through the
swampland together with continued advances in string compactifications
would enable us to better understand the rigid pattern of particle
spectra found in string theory.

\section{Connections to other areas of research}

\subsection{Mathematics}
 \label{sec:math}

Since the early days of string theory, there has been a constant flux
of new ideas in both directions between physicists studying string
theory and a wide range of branches of mathematics.  This synergy
between the fields has continued unabated, and perhaps even increased,
in recent years.  One particularly strong set of such connections
relates to the use of geometry in string compactifications.  In
particular, in recent years there have been exciting developments
generating new connections and insights for both math and physics
related to special holonomy (``$G_2$'' manifolds) relevant for
compactifications of 11-dimensional M-theory to 4D, and to the
algebraic geometry of elliptically fibered Calabi-Yau threefolds and
fourfolds, relevant for compactification of F-theory to 6D and 4D.
These developments promise to provide powerful tools for better
understanding F-theory and M-theory constructions of Standard
Model-like string vacua.

In the context of F-theory, there has been significant progress in
understanding many aspects of elliptically fibered Calabi-Yau threefolds and
fourfolds.  Evidence suggests that most known Calabi-Yau threefolds
and fourfolds are in fact elliptically fibered
\cite{Gray:2014fla,Anderson:2017aux,Huang:2018esr,Huang:2019pne}.   
New mathematical
results on the finiteness of topological equivalence classes of
Calabi-Yau fourfolds \cite{DiCerboSvaldi}, expanding the earlier results on Calabi-Yau
threefolds \cite{Grassi,Gross-finite},
give insight into the global structure of
string vacua, in general, and F-theory vacua, in particular.  Progress has been made  on
generalizing the Kodaira classification of codimension one
singularities in elliptic fibrations, which matches beautifully with
the nonperturbative physics of non-Abelian gauge groups, by
understanding higher-codimension singularities encoding matter and
Yukawa couplings (for some examples see, e.g.,
\cite{Katz:1996xe,Arras:2016evy,Esole:2017kyr,Klevers:2017aku}).  
Difficult mathematical problems associated
with the Mordell-Weil and Tate-Shafarevich/Weil-Ch\^atelet groups of
an elliptically fibered Calabi-Yau are
connected with physics of continuous and discrete Abelian gauge
symmetries in F-theory, and progress is being made in understanding
these structures with mutual benefit to mathematicians and
physicists. For a comprehensive review on these developments, see \cite{Cvetic:2018bni}.  There are a number of open questions in these areas that
are promising for progress in the coming years and will shed light on
important questions both in physics and math.

M-theory compactifications on seven-manifolds with $G_2$ holonomy also
give rise to large ensembles of $4d$ $\mathcal{N}=1$ vacua, where
singularities at codimension $4$, $6$, and $7$, encode the structure
of gauge groups, non-chiral matter, and chiral matter, respectively;
see \cite{Acharya:2004qe} for an early review.  More recently,
constructions of so-called twisted connected sums
\cite{Kovalev:2001zr} have given rise to millions of $G_2$ manifolds
\cite{Corti:2012kd}, which have been the subject of numerous physics
studies in compact twisted connected sums \cite{Halverson:2014tya,
  Halverson:2015vta,Braun:2016igl,daCGuio:2017ifs,Braun:2017ryx,Braun:2017uku,
  Braun:2017csz, Braun:2018fdp}, and related Higgs bundle
constructions
\cite{Pantev:2009de,Braun:2018vhk,Barbosa:2019bgh,Hubner:2020yde,Cvetic:2020piw}. For
some recent progress on singular non-compact $G_2$ constructions and
studies on gauge dynamics there, see
\cite{Acharya:2020vmg,Cvetic:2022imb}.  In general, though, much less is known about compact
(singular) $G_2$ manifolds than Calabi-Yau compactifications. This is
in part due to the fact that there is no analog of Yau's theorem,
i.e., no simple topological check that guarantees the existence of a
Ricci-flat metric with appropriate holonomy.

One set of questions that is common to the challenges of the $G_2$
special holonomy research and the elliptically fibered Calabi-Yau F-theory
research is the challenge of understanding better how mathematical
structures such as intersection theory operate in singular spaces.
This is only partially understood by mathematicians and only in
limited domains but is crucial to understanding the physics of string
compactifications in both these areas.  In particular, in $G_2$
holonomy manifolds and elliptically fibered Calabi-Yau varieties, singularities
are essential for the physics of non-Abelian gauge theories, matter
fields, Yukawa couplings and other important aspects of the theory.  The primary
approach taken to F-theory currently is to view it as a limit of a
compactification of M-theory on a smooth Calabi-Yau, but a more
intrinsic definition is given by IIB string theory, which is
characterized by singular elliptic fibrations associated with 7-brane
configurations. Finding a direct description of the physics of
F-theory in terms of the singular geometries may be a crucial step to
a better general class of tools for describing details of
compactification; such a description, however, requires understanding
intersection theory on singular spaces.  Recent work
\cite{Grassi:2018wfy,Grassi:2021ptc,Jefferson:2021bid} shows how some
of these features must be properties of the singular spaces, providing
examples of how a proper mathematical theory of these singularities
should work; providing a systematic mathematically sound methodology
for analyzing these spaces is an important challenge for the future.

A simple illustration demonstrates the importance of an appropriate homology and
intersection theory on singular compactification spaces. Since gauge
sectors arise at codimension $4$ in $G_2$ compactifications, and two
codimension $4$ cycles do not intersect in a $7$-manifold, one might
conclude that distinct gauge sector loci cannot have jointly charged
matter in $G_2$ compactifications. However, this directly contradicts
the existence of uplifts of IIA models with intersecting D6-branes
\cite{Cvetic:2001kk}, and also local $G_2$ constructions of unfolding chiral matter \cite{Acharya:2001gy}. The error is
that the non-intersection of three-cycles holds in
(smooth) $7$-manifolds, but
not in seven-dimension singular spaces. One way to correct this is to determine
an appropriate homology and intersection theory in the
presence of singularities, perhaps intersection homology
\cite{GORESKY1980135} of Goresky and Macpherson, which corrects the
failure of Poincar\'e duality on singular spaces.

\subsection{Machine Learning and Computational Complexity}
\label{sec:ml}

As our understanding of the theoretical framework for string
compactifications increases, the problems in identifying the desired
vacua and computing their characteristics become increasingly
well-defined.  Many of these problems are computationally challenging
and involve nonperturbative physics or exponentially large search
spaces, and will likely require sophisticated computational
approaches, just as many of the detailed features of strongly coupled
quantum field theories like QCD are currently best understood through
lattice gauge computations.  There have been efforts to use
computational approaches to analyze a number of problems related to
string compactifications, such as for the computation of exact metrics
on Calabi-Yau compactification spaces, for which no analytic solution
is known
\cite{Headrick:2005ch,Donaldsonnumerical,Douglas:2006rr,Braun:2007sn,Anderson:2010ke,Kachru:2020tat}.
In this section we focus on the use of modern methods of machine
learning and associated methodologies for approaching some of the
computational difficult challenges in string compactification.

The string landscape is vast, and our knowledge of it has grown
significantly in the last decade. Evidence for exponentially large
numbers of $4$d string vacua was already given in 1987
\cite{Lerche:1986cx}, in the context of chiral heterotic
models. However, the possibility received increased attention with
work by Bousso and Polchinski \cite{Bousso:2000xa}, which provided a
string theoretic mechanism for realizing Weinberg's anthropic solution
to the cosmological constant problem \cite{Weinberg:1987dv}. Famously,
this led to an estimate of $10^{500}$ flux vacua in weakly coupled
type IIB compactifications (reviewed in \cite{DenefLesHouches}), an
estimate that recently ballooned to
$10^{272,000}$~\cite{Taylor:2015xtz} by moving outside of the weakly
coupled regime to F-theory. A more recent development is that the
number of string geometries is also exponentially large. Though
Kreuzer and Skarke's classification relating four-dimensional
reflexive polytopes and Calabi-Yau threefolds yields only a strict
lower bound of $O(30,000)$ Calabi-Yau threefolds (from distinct Hodge
numbers), this was always expected to be a significant undercount due to
triangulated polytope combinatorics, and the number of Calabi-Yau
fourfolds is dramatically larger, with a recently studied subset
giving over 500 million distinct Hodge numbers for fourfolds
\cite{Scholler:2018apc}.
A related setup in F-theory
instead counts bases of elliptically fibered Calabi-Yau threefolds
\cite{Morrison:2012js,Taylor:2015isa} and fourfolds
\cite{Anderson:2014gla,Halverson:2015jua,Taylor:2015ppa}.  Compared to
roughly 65,000 distinct toric bases for threefolds, this approach provides a
strict lower bound of $O(10^{755})$ F-theory geometries for fourfold
bases \cite{Halverson:2017ffz}, which is also a vast undercount due to
imposing a sufficient but not necessary condition; Monte Carlo
estimates \cite{Taylor:2017yqr} suggest there are $O(10^{3000})$
geometries with the relaxed condition. Despite these large numbers of
elliptically fibered Calabi-Yau fourfolds, a result appeared around
the same time proving that the total number is finite
\cite{DiCerboSvaldi}.  However, the string landscape is not only vast,
it is also unwieldy: computationally complex problems abound. For
instance, the search for small cosmological constants in simplified
idealized models is already NP-hard \cite{Denef:2006ad}, although the
success in finding explicit solutions with small cosmological constant
in F-theory \cite{Demirtas:2021nlu} suggests that the structure of the
landscape may enable efficient solutions to this problem.
Furthermore, \cite{Halverson:2018cio} computing effective potentials
often requires solving instances of NP-hard problems, and the search
for local minima is itself (co)-NP-hard. Additionally, the appearance
of both explicit diophantine equations in string theory (from both
index theorems, e.g., and diophantine encodings of decision problems)
brings \emph{undecidability} into the game, by the negative solution
to Hilbert's tenth problem \cite{Cvetic:2010ky}. 
% An interesting recent
% avenue is to consider dynamical cosmological measures that are motived
% by ideas from complexity theory \cite{Denef:2017cxt,Khoury:2019yoo}.
There are a number of potential mechanisms for avoiding complexity
issues, however. For instance, landscape structure may aid in solving
complex problems such as Diophantines \cite{Halverson:2019vmd};
fast-enough algorithms may exist for system sizes of interest
\cite{Bao:2017thx}; some NP-hard problems have fully polynomial time approximation schemes, which allow for polynomial time solution if small errors are allowed; and complexity considerations can change by allowing for stochastic or quantum computers.

Taken together, the enormity and complexity of the landscape
motivates the use of modern techniques from computer science. Much
of the focus has been on machine learning, beginning with
\cite{He:2017aed,Ruehle:2017mzq,Krefl:2017yox,Carifio:2017bov}, which
utilized supervised learning (both with and without neural networks)
and led to machine-assisted theorems via conjecture generation. Other
notable areas include persistent homology, which can detect cycles and
voids in the landscape \cite{cirafici2016persistent,Cole:2018emh};
network science, which can be used to model tunneling transitions and
dynamical measures \cite{Carifio:2017nyb}; fast SAT and SMT solvers \cite{Faraggi:2021mws, Faraggi:2022hut}, which can solve string constraints by mapping them to famous problems in computer science, SAT and SMT; and genetic algorithms
\cite{Blaback:2013ht,Blaback:2013fca,Abel:2014xta,Ruehle:2017mzq,Cole:2019enn,AbdusSalam:2020ywo,Bena:2021wyr,
  Loges:2021hvn}, which use ideas from evolution to search for
solutions, e.g., for string vacua satisfying various properties.

A number of deep learning techniques have been utilized to study the
string landscape and associated mathematical data. Due to the enormity
of the landscape, we organize the discussion here according to the
type of question under consideration, and the associated machine
learning technique.  \emph{Prediction} is the domain of supervised
learning, whereby a neural network or simpler algorithm is trained to
predict outputs given inputs. String theoretic applications of
supervised learning
(e.g. \cite{He:2017aed,Ruehle:2017mzq,Krefl:2017yox,Carifio:2017bov,Liu:2017dzi,Wang:2018rkk,Altman:2018zlc,Jinno:2018dek,Bull:2018uow,Rudelius:2018yqi,Jejjala:2019kio}),
often predict physical features such as gauge group and Hodge numbers
(which includes number of axions) given core geometric, D-brane, or
flux data as input. Notably, conjectures may be generated by bringing
the human into the loop: though supervised learning has intrinstic
error, it can also learn correlations that can be understood by
humans, especially when simpler techniques are used, that can lead to
conjectures and even theorems \cite{Carifio:2017bov,
  Brodie:2019dfx,Bies:2020gvf}.  Though supervised learning was the
focus of early work, new directions quickly arose. \emph{Search} for
vacua with particular properties
\cite{Halverson:2019tkf,Larfors:2020ugo,Krippendorf:2021uxu,Constantin:2021for}
may be carried out with deep Reinforcement Learning (RL), by which a
trained neural network represents a learned policy function that
chooses intelligent actions, given the state of the system; RL is the
type of deep learning utilized by DeepMind in its famous works on Go
and Chess \cite{Silver2017}. For instance, the state could be an intermediate
stage of constructing a string compactification with D-branes, where
intelligent actions would push the system towards global consistency
and interesting phenomenology. Genetic algorithms, an optimization
technique that does not utilize a neural network, also regularly lead
to good search results; see
\cite{Cole:2021nnt,Abel:2021ddu,Abel:2021rrj} for comparisons.  Deep
generative models are trained to sample from a desired probability
distribution, and may be utilized to \emph{simulate} SUSY EFT data
\cite{Erbin:2018csv} or string data \cite{Halverson:2020opj}. Finally,
a major recent and very notable development is in
\emph{self-generative learning}, where the neural network itself is
trained to represent a function of interest, usually a solution to a
PDE, such as a Calabi-Yau metric \cite{Anderson:2020hux,
  Douglas:2020hpv, Jejjala:2020wcc}, which was recently extended to
general Kreuzer-Skarke Calabi-Yau threefolds
\cite{Larfors:2021pbb}; see also \cite{Ashmore:2019wzb}. Interestingly, these closely resemble
techniques used to learn ground states of quantum many-body systems
\cite{2017Sci...355..602C}. For a recent review, see
\cite{RUEHLE20201}.

While it is not yet clear how far machine learning and related
methodologies can go in addressing these difficult string vacuum questions,
these developments lay the groundwork for a deepened understanding of
the landscape that may be obtained via machine
learning. Self-generative learning, including in the context of
Calabi-Yau metrics, opens the door to the study of non-holomorphic
data such as non-BPS charged particles and Kaluza-Klein modes that are
difficult to study with traditional techniques. These same techniques
can push the envelope in pure mathematics, for instance in studies of
$G_2$ manifolds and singular spaces with numerical metrics. Search, as
offered by reinforcement learning and genetic algorithms, provides new
opportunity to understand what is possible in string theory. %Excellent
Applications include the search for vacua with potential for realistic
particle physics and cosmology, as well as other observables typical
of those vacua. Specifically, these techniques could be utilized to
efficiently search for vacua with small positive cosmological
constants. Finally, generative models provide a means of sampling from
desired distributions, which will be essential to making statistical
predictions
with some 
measures; see \cite{Halverson:2020opj}.

\section{Outlook}

As we have reviewed here, connecting the top-down framework of string
theory with the bottom-up observational data of particle physics is an
active and vibrant enterprise and is an important central component of
completing our understanding of the Universe.  In the last decade,
substantial progress has been made in framing and beginning to address
the questions listed in the introduction: How precisely can string
theory match observed particle physics? What is ruled out from string
theory?  What are typical features of string vacuum solutions and what
are the consequences for particle physics?  The nonperturbative
approach of F-theory and other developments in string compactification
have given an increasingly global perspective on the set of string
solutions from which these questions can be addressed.  There are now
substantial classes of top-down string constructions of vacua that
contain the Standard Model gauge group and the three family chiral
matter content, and some of these appear to involve fairly minimal
fine tuning.  Certain features, such as axions and strongly coupled
hidden sectors arise ubiquitously in string vacuum constructions, and
suggest natural dark matter candidates as well as potential avenues
for applying cosmological constraints.  Constraints on low-energy
theories from string theory promise to shed insight on long-standing
questions such as the hierarchy problem.

Many key questions, however, remain unanswered.
The lack of experimental observation of low-energy supersymmetry
sharpens the questions about the physics of SUSY breaking and the
nature of non-supersymmetric solutions in string theory.
The observed small positive cosmological constant sharpens the
challenge of understanding de Sitter and non-supersymmetric string
vacua. The landscape is large and while naive counting suggests that
certain features may dominate,
the measure problem is open, which is crucial for making any kind of
precise statistical statement regarding string vacua.

This general research area, which aims at connections between
UV-complete quantum gravity theories and the observed Standard Model
of particle physics, promises to be a very exciting and dynamic area of
activity in the coming decade, and  brings together the 
research efforts of formal theorists  with the large community of
particle physicists working closer to experiment.  Supporting this
effort should be a crucial part of DOE high-energy priorities through
the 2020s and 2030s.

\section*{Acknowledgements}
The work of M.C.is supported by DOE Award DE-SC0013528Y, the Simons Foundation Collaboration grant \#724069 on ``Special Holonomy in Geometry, Analysis and Physics'', the Slovenian Research Agency No.~P1-0306 and the Fay R.~and Eugene L.~Langberg Chair funds.
The work of GS is supported in part by the DOE grant DE-SC0017647. 
The work of WT is supported in part by the DOE through grant DE-SC00012567. 
The work of J.H. is supported by NSF
CAREER grant PHY-1848089.

\bibliographystyle{utphys}
\bibliography{snowmass_refs.bib}

\providecommand{\href}[2]{#2}\begingroup\raggedright\begin{thebibliography}{100}

\bibitem{Gross:1984dd}
D.~J. Gross, J.~A. Harvey, E.~J. Martinec and R.~Rohm,  {\em {The Heterotic
  String}}, Phys. Rev. Lett. {\bf 54} (1985) 502--505.

\bibitem{Candelas:1985en}
P.~Candelas, G.~T. Horowitz, A.~Strominger and E.~Witten,  {\em {Vacuum
  Configurations for Superstrings}}, Nucl. Phys. B {\bf 258} (1985) 46--74.

\bibitem{Flauger:2022hie}
R.~Flauger, V.~Gorbenko, A.~Joyce, L.~McAllister, G.~Shiu and E.~Silverstein,
  {\em {Snowmass White Paper: Cosmology at the Theory Frontier}}, in {\em {2022
  Snowmass Summer Study}}.
\newblock 3, 2022.
\newblock \href{http://www.arXiv.org/abs/2203.07629}{{\tt 2203.07629}}.

\bibitem{VafaF-theory}
C.~Vafa,  {\em {Evidence for F theory}}, Nucl. Phys. {\bf B469} (1996) 403--418
[\href{http://www.arXiv.org/abs/hep-th/9602022}{{\tt hep-th/9602022}}].
%%CITATION = HEP-TH/9602022;%%.

\bibitem{MorrisonVafaI}
D.~R. Morrison and C.~Vafa,  {\em {Compactifications of F theory on Calabi--Yau
  threefolds --- I}}, Nucl. Phys. {\bf B473} (1996) 74--92
[\href{http://www.arXiv.org/abs/hep-th/9602114}{{\tt hep-th/9602114}}].
%%CITATION = HEP-TH/9602114;%%.

\bibitem{MorrisonVafaII}
D.~R. Morrison and C.~Vafa,  {\em {Compactifications of F theory on Calabi--Yau
  threefolds --- II}}, Nucl. Phys. {\bf B476} (1996) 437--469
[\href{http://www.arXiv.org/abs/hep-th/9603161}{{\tt hep-th/9603161}}].
%%CITATION = HEP-TH/9603161;%%.

\bibitem{Greene:1986ar}
B.~R. Greene, K.~H. Kirklin, P.~J. Miron and G.~G. Ross,  {\em {A Superstring
  Inspired Standard Model}}, Phys. Lett. {\bf B180} (1986)
69.
%%CITATION = PHLTA,B180,69;%%.

\bibitem{Braun:2005ux}
V.~Braun, Y.-H. He, B.~A. Ovrut and T.~Pantev,  {\em {A Heterotic standard
  model}}, Phys. Lett. B {\bf 618} (2005) 252--258
  [\href{http://www.arXiv.org/abs/hep-th/0501070}{{\tt hep-th/0501070}}].

\bibitem{Bouchard:2005ag}
V.~Bouchard and R.~Donagi,  {\em {An SU(5) heterotic standard model}}, Phys.
  Lett. {\bf B633} (2006) 783--791
[\href{http://www.arXiv.org/abs/hep-th/0512149}{{\tt hep-th/0512149}}].
%\%CITATION = HEP-TH/0512149;\%\%.

\bibitem{Bouchard:2006dn}
V.~Bouchard, M.~Cveti{\v c} and R.~Donagi,  {\em {Tri-linear couplings in an
  heterotic minimal supersymmetric standard model}}, Nucl. Phys. B {\bf 745}
  (2006) 62--83 [\href{http://www.arXiv.org/abs/hep-th/0602096}{{\tt
  hep-th/0602096}}].

\bibitem{Anderson:2009mh}
L.~B. Anderson, J.~Gray, Y.-H. He and A.~Lukas,  {\em {Exploring Positive Monad
  Bundles And A New Heterotic Standard Model}}, JHEP {\bf 02} (2010) 054
[\href{http://www.arXiv.org/abs/0911.1569}{{\tt 0911.1569}}].
%%CITATION = ARXIV:0911.1569;%%.

\bibitem{Anderson:2011ns}
L.~B. Anderson, J.~Gray, A.~Lukas and E.~Palti,  {\em {Two Hundred Heterotic
  Standard Models on Smooth Calabi-Yau Threefolds}}, Phys. Rev. D {\bf 84}
  (2011) 106005 [\href{http://www.arXiv.org/abs/1106.4804}{{\tt 1106.4804}}].

\bibitem{Anderson:2012yf}
L.~B. Anderson, J.~Gray, A.~Lukas and E.~Palti,  {\em {Heterotic Line Bundle
  Standard Models}}, JHEP {\bf 06} (2012) 113
[\href{http://www.arXiv.org/abs/1202.1757}{{\tt 1202.1757}}].
%%CITATION = ARXIV:1202.1757;%%.

\bibitem{Lebedev:2007hv}
O.~Lebedev, H.~P. Nilles, S.~Raby, S.~Ramos-Sanchez, M.~Ratz, P.~K.~S.
  Vaudrevange and A.~Wingerter,  {\em {The Heterotic Road to the MSSM with R
  parity}}, Phys. Rev. D {\bf 77} (2008) 046013
  [\href{http://www.arXiv.org/abs/0708.2691}{{\tt 0708.2691}}].

\bibitem{Lebedev:2006kn}
O.~Lebedev, H.~P. Nilles, S.~Raby, S.~Ramos-Sanchez, M.~Ratz, P.~K.~S.
  Vaudrevange and A.~Wingerter,  {\em {A Mini-landscape of exact MSSM spectra
  in heterotic orbifolds}}, Phys. Lett. B {\bf 645} (2007) 88--94
  [\href{http://www.arXiv.org/abs/hep-th/0611095}{{\tt hep-th/0611095}}].

\bibitem{Faraggi:1989ka}
A.~E. Faraggi, D.~V. Nanopoulos and K.-j. Yuan,  {\em {A Standard Like Model in
  the 4D Free Fermionic String Formulation}}, Nucl. Phys. B {\bf 335} (1990)
  347--362.

\bibitem{Faraggi:1991jr}
A.~E. Faraggi,  {\em {A New standard - like model in the four-dimensional free
  fermionic string formulation}}, Phys. Lett. B {\bf 278} (1992) 131--139.

\bibitem{Anderson:2013xka}
L.~B. Anderson, A.~Constantin, J.~Gray, A.~Lukas and E.~Palti,  {\em {A
  Comprehensive Scan for Heterotic SU(5) GUT models}}, JHEP {\bf 01} (2014) 047
  [\href{http://www.arXiv.org/abs/1307.4787}{{\tt 1307.4787}}].

\bibitem{Polchinski:1995mt}
J.~Polchinski,  {\em {Dirichlet Branes and Ramond-Ramond charges}}, Phys. Rev.
  Lett. {\bf 75} (1995) 4724--4727
  [\href{http://www.arXiv.org/abs/hep-th/9510017}{{\tt hep-th/9510017}}].

\bibitem{Shiu:1998pa}
G.~Shiu and S.~H.~H. Tye,  {\em {TeV scale superstring and extra dimensions}},
  Phys. Rev. D {\bf 58} (1998) 106007
  [\href{http://www.arXiv.org/abs/hep-th/9805157}{{\tt hep-th/9805157}}].

\bibitem{Aldazabal:2000sa}
G.~Aldazabal, L.~E. Ibanez, F.~Quevedo and A.~M. Uranga,  {\em {D-branes at
  singularities: A Bottom up approach to the string embedding of the standard
  model}}, JHEP {\bf 08} (2000) 002
  [\href{http://www.arXiv.org/abs/hep-th/0005067}{{\tt hep-th/0005067}}].

\bibitem{Cvetic:2000st}
M.~Cveti\v{c}, A.~M. Uranga and J.~Wang,  {\em {Discrete Wilson lines in N=1 D
  = 4 type IIB orientifolds: A Systematic exploration for Z(6) orientifold}},
  Nucl. Phys. B {\bf 595} (2001) 63--92
  [\href{http://www.arXiv.org/abs/hep-th/0010091}{{\tt hep-th/0010091}}].

\bibitem{Berenstein:2001nk}
D.~Berenstein, V.~Jejjala and R.~G. Leigh,  {\em {The Standard model on a
  D-brane}}, Phys. Rev. Lett. {\bf 88} (2002) 071602
  [\href{http://www.arXiv.org/abs/hep-ph/0105042}{{\tt hep-ph/0105042}}].

\bibitem{Verlinde:2005jr}
H.~Verlinde and M.~Wijnholt,  {\em {Building the standard model on a
  D3-brane}}, JHEP {\bf 01} (2007) 106
  [\href{http://www.arXiv.org/abs/hep-th/0508089}{{\tt hep-th/0508089}}].

\bibitem{Berkooz:1996km}
M.~Berkooz, M.~R. Douglas and R.~G. Leigh,  {\em {Branes intersecting at
  angles}}, Nucl. Phys. {\bf B480} (1996) 265--278
[\href{http://www.arXiv.org/abs/hep-th/9606139}{{\tt hep-th/9606139}}].
%%CITATION = HEP-TH/9606139;%%.

\bibitem{Aldazabal:2000dg}
G.~Aldazabal, S.~Franco, L.~E. Ibanez, R.~Rabadan and A.~M. Uranga,  {\em {D =
  4 chiral string compactifications from intersecting branes}}, J. Math. Phys.
  {\bf 42} (2001) 3103--3126
[\href{http://www.arXiv.org/abs/hep-th/0011073}{{\tt hep-th/0011073}}].
%%CITATION = HEP-TH/0011073;%%.

\bibitem{Aldazabal:2000cn}
G.~Aldazabal, S.~Franco, L.~E. Ibanez, R.~Rabadan and A.~M. Uranga,  {\em
  {Intersecting brane worlds}}, JHEP {\bf 02} (2001) 047
[\href{http://www.arXiv.org/abs/hep-ph/0011132}{{\tt hep-ph/0011132}}].
%%CITATION = HEP-PH/0011132;%%.

\bibitem{Ibanez:2001nd}
L.~E. Ibanez, F.~Marchesano and R.~Rabadan,  {\em {Getting just the standard
  model at intersecting branes}}, JHEP {\bf 11} (2001) 002
[\href{http://www.arXiv.org/abs/hep-th/0105155}{{\tt hep-th/0105155}}].
%%CITATION = HEP-TH/0105155;%%.

\bibitem{Blumenhagen:2001te}
R.~Blumenhagen, B.~Kors, D.~L{\"u}st and T.~Ott,  {\em {The standard model from
  stable intersecting brane world orbifolds}}, Nucl. Phys. {\bf B616} (2001)
  3--33
[\href{http://www.arXiv.org/abs/hep-th/0107138}{{\tt hep-th/0107138}}].
%%CITATION = HEP-TH/0107138;%%.

\bibitem{Cvetic:2001tj}
M.~Cveti{\v c}, G.~Shiu and A.~M. Uranga,  {\em {Three family supersymmetric
  standard - like models from intersecting brane worlds}}, Phys. Rev. Lett.
  {\bf 87} (2001) 201801
[\href{http://www.arXiv.org/abs/hep-th/0107143}{{\tt hep-th/0107143}}].
%%CITATION = HEP-TH/0107143;%%.

\bibitem{Cvetic:2001nr}
M.~Cveti{\v c}, G.~Shiu and A.~M. Uranga,  {\em {Chiral four-dimensional N=1
  supersymmetric type 2A orientifolds from intersecting D6 branes}}, Nucl.
  Phys. {\bf B615} (2001) 3--32
[\href{http://www.arXiv.org/abs/hep-th/0107166}{{\tt hep-th/0107166}}].
%%CITATION = HEP-TH/0107166;%%.

\bibitem{Blumenhagen:2005mu}
R.~Blumenhagen, M.~Cveti{\v c}, P.~Langacker and G.~Shiu,  {\em {Toward
  realistic intersecting D-brane models}}, Ann. Rev. Nucl. Part. Sci. {\bf 55}
  (2005) 71--139
[\href{http://www.arXiv.org/abs/hep-th/0502005}{{\tt hep-th/0502005}}].
%%CITATION = HEP-TH/0502005;%%.

\bibitem{Acharya:2008zi}
B.~S. Acharya, K.~Bobkov, G.~L. Kane, J.~Shao and P.~Kumar,  {\em {The
  G(2)-MSSM: An M Theory motivated model of Particle Physics}}, Phys. Rev. D
  {\bf 78} (2008) 065038 [\href{http://www.arXiv.org/abs/0801.0478}{{\tt
  0801.0478}}].

\bibitem{Vafa:1996xn}
C.~Vafa,  {\em {Evidence for F theory}}, Nucl. Phys. {\bf B469} (1996) 403--418
[\href{http://www.arXiv.org/abs/hep-th/9602022}{{\tt hep-th/9602022}}].
%%CITATION = HEP-TH/9602022;%%.

\bibitem{oai:arXiv.org:hep-th/9602114}
D.~R. Morrison and C.~Vafa,  {\em {Compactifications of F theory on Calabi-Yau
  threefolds. 1}}, Nucl.Phys. {\bf B473} (1996) 74--92
[\href{http://www.arXiv.org/abs/hep-th/9602114}{{\tt hep-th/9602114}}].
%\%CITATION = HEP-TH/9602114;\%\%.

\bibitem{oai:arXiv.org:hep-th/9603161}
D.~R. Morrison and C.~Vafa,  {\em {Compactifications of F theory on Calabi-Yau
  threefolds. 2.}}, Nucl.Phys. {\bf B476} (1996) 437--469
[\href{http://www.arXiv.org/abs/hep-th/9603161}{{\tt hep-th/9603161}}].
%\%CITATION = HEP-TH/9603161;\%\%.

\bibitem{DonagiWijnholtGUTs}
R.~Donagi and M.~Wijnholt,  {\em {Breaking GUT Groups in F-Theory}}, Adv.
  Theor. Math. Phys. {\bf 15} (2011), no.~6, 1523--1603
[\href{http://www.arXiv.org/abs/0808.2223}{{\tt 0808.2223}}].
%%CITATION = ARXIV:0808.2223;%%.

\bibitem{BeasleyHeckmanVafaI}
C.~Beasley, J.~J. Heckman and C.~Vafa,  {\em {GUTs and Exceptional Branes in
  F-theory - I}}, JHEP {\bf 01} (2009) 058
[\href{http://www.arXiv.org/abs/0802.3391}{{\tt 0802.3391}}].
%%CITATION = ARXIV:0802.3391;%%.

\bibitem{BeasleyHeckmanVafaII}
C.~Beasley, J.~J. Heckman and C.~Vafa,  {\em {GUTs and Exceptional Branes in
  F-theory - II: Experimental Predictions}}, JHEP {\bf 01} (2009) 059
[\href{http://www.arXiv.org/abs/0806.0102}{{\tt 0806.0102}}].
%%CITATION = ARXIV:0806.0102;%%.

\bibitem{DonagiWijnholtModelBuilding}
R.~Donagi and M.~Wijnholt,  {\em {Model Building with F-Theory}}, Adv. Theor.
  Math. Phys. {\bf 15} (2011), no.~5, 1237--1317
[\href{http://www.arXiv.org/abs/0802.2969}{{\tt 0802.2969}}].
%%CITATION = ARXIV:0802.2969;%%.

\bibitem{GrassiHalversonShanesonTaylor}
A.~Grassi, J.~Halverson, J.~Shaneson and W.~Taylor,  {\em {Non-Higgsable QCD
  and the Standard Model Spectrum in F-theory}}, JHEP {\bf 01} (2015) 086
[\href{http://www.arXiv.org/abs/1409.8295}{{\tt 1409.8295}}].
%%CITATION = ARXIV:1409.8295;%%.

\bibitem{Klevers:2014bqa}
D.~Klevers, D.~K. Mayorga~Pe{\~ n}a, P.-K. Oehlmann, H.~Piragua and J.~Reuter,
  {\em {F-Theory on all Toric Hypersurface Fibrations and its Higgs Branches}},
  JHEP {\bf 01} (2015) 142
[\href{http://www.arXiv.org/abs/1408.4808}{{\tt 1408.4808}}].
%%CITATION = ARXIV:1408.4808;%%.

\bibitem{Cvetic:2015txa}
M.~Cveti{\v c}, D.~Klevers, D.~K.~M. Pe{\~ n}a, P.-K. Oehlmann and J.~Reuter,
  {\em {Three-Family Particle Physics Models from Global F-theory
  Compactifications}}, JHEP {\bf 08} (2015) 087
[\href{http://www.arXiv.org/abs/1503.02068}{{\tt 1503.02068}}].
%%CITATION = ARXIV:1503.02068;%%.

\bibitem{Taylor:2019wnm}
W.~Taylor and A.~P. Turner,  {\em {Generic Construction of the Standard Model
  Gauge Group and Matter Representations in F-theory}}, Fortsch. Phys. {\bf 68}
  (2020), no.~5, 2000009 [\href{http://www.arXiv.org/abs/1906.11092}{{\tt
  1906.11092}}].

\bibitem{Raghuram:2019efb}
N.~Raghuram, W.~Taylor and A.~P. Turner,  {\em {General F-theory models with
  tuned $(\operatorname{SU}(3) \times \operatorname{SU}(2) \times
  \operatorname{U}(1)) / \mathbb{Z}_6$ symmetry}}, JHEP {\bf 04} (2020) 008
  [\href{http://www.arXiv.org/abs/1912.10991}{{\tt 1912.10991}}].

\bibitem{Li:2021eyn}
S.~Y. Li and W.~Taylor,  {\em {Natural F-theory constructions of Standard Model
  structure from $E_7$ flux breaking}},
  \href{http://www.arXiv.org/abs/2112.03947}{{\tt 2112.03947}}.

\bibitem{Marsano:2010ix}
J.~Marsano, N.~Saulina and S.~Schafer-Nameki,  {\em {A Note on G-Fluxes for
  F-theory Model Building}}, JHEP {\bf 11} (2010) 088
  [\href{http://www.arXiv.org/abs/1006.0483}{{\tt 1006.0483}}].

\bibitem{Braun:2011zm}
A.~P. Braun, A.~Collinucci and R.~Valandro,  {\em {G-flux in F-theory and
  algebraic cycles}}, Nucl. Phys. {\bf B856} (2012) 129--179
[\href{http://www.arXiv.org/abs/1107.5337}{{\tt 1107.5337}}].
%\%CITATION = 1107.5337;\%\%.

\bibitem{Marsano:2011hv}
J.~Marsano and S.~Schafer-Nameki,  {\em {Yukawas, G-flux, and Spectral Covers
  from Resolved Calabi-Yau's}}, JHEP {\bf 11} (2011) 098
[\href{http://www.arXiv.org/abs/1108.1794}{{\tt 1108.1794}}].
%%CITATION = ARXIV:1108.1794;%%.

\bibitem{oai:arXiv.org:1111.1232}
T.~W. Grimm and H.~Hayashi,  {\em {F-theory fluxes, Chirality and Chern-Simons
  theories}}, JHEP {\bf 1203} (2012) 027
[\href{http://www.arXiv.org/abs/1111.1232}{{\tt 1111.1232}}].
%\%CITATION = ARXIV:1111.1232;\%\%.

\bibitem{Krause:2012yh}
S.~Krause, C.~Mayrhofer and T.~Weigand,  {\em {Gauge Fluxes in F-theory and
  Type IIB Orientifolds}}, JHEP {\bf 08} (2012) 119
[\href{http://www.arXiv.org/abs/1202.3138}{{\tt 1202.3138}}].
%%CITATION = ARXIV:1202.3138;%%.

\bibitem{Braun:2013nqa}
V.~Braun, T.~W. Grimm and J.~Keitel,  {\em {Geometric Engineering in Toric
  F-Theory and GUTs with U(1) Gauge Factors}}, JHEP {\bf 12} (2013) 069
[\href{http://www.arXiv.org/abs/1306.0577}{{\tt 1306.0577}}].
%%CITATION = ARXIV:1306.0577;%%.

\bibitem{Cvetic:2013uta}
M.~Cveti{\v c}, A.~Grassi, D.~Klevers and H.~Piragua,  {\em {Chiral
  Four-Dimensional F-Theory Compactifications With SU(5) and Multiple
  U(1)-Factors}}, JHEP {\bf 04} (2014) 010
[\href{http://www.arXiv.org/abs/1306.3987}{{\tt 1306.3987}}].
%%CITATION = ARXIV:1306.3987;%%.

\bibitem{Lin:2015qsa}
L.~Lin, C.~Mayrhofer, O.~Till and T.~Weigand,  {\em {Fluxes in F-theory
  Compactifications on Genus-One Fibrations}}, JHEP {\bf 01} (2016) 098
[\href{http://www.arXiv.org/abs/1508.00162}{{\tt 1508.00162}}].
%%CITATION = ARXIV:1508.00162;%%.

\bibitem{Lin:2016vus}
L.~Lin and T.~Weigand,  {\em {G4-flux and standard model vacua in F-theory}},
  Nucl. Phys. {\bf B913} (2016) 209--247
[\href{http://www.arXiv.org/abs/1604.04292}{{\tt 1604.04292}}].
%%CITATION = ARXIV:1604.04292;%%.

\bibitem{Jefferson:2021bid}
P.~Jefferson, W.~Taylor and A.~P. Turner,  {\em {Chiral matter multiplicities
  and resolution-independent structure in 4D F-theory models}},
  \href{http://www.arXiv.org/abs/2108.07810}{{\tt 2108.07810}}.

\bibitem{Krause:2011xj}
S.~Krause, C.~Mayrhofer and T.~Weigand,  {\em {$G_4$ flux, chiral matter and
  singularity resolution in F-theory compactifications}}, Nucl. Phys. {\bf
  B858} (2012) 1--47
[\href{http://www.arXiv.org/abs/1109.3454}{{\tt 1109.3454}}].
%%CITATION = ARXIV:1109.3454;%%.

\bibitem{Cvetic:2018ryq}
M.~Cveti\v{c}, L.~Lin, M.~Liu and P.-K. Oehlmann,  {\em {An F-theory
  Realization of the Chiral MSSM with $\mathbb{Z}_2$-Parity}}, JHEP {\bf 09}
  (2018) 089
[\href{http://www.arXiv.org/abs/1807.01320}{{\tt 1807.01320}}].
%%CITATION = ARXIV:1807.01320;%%.

\bibitem{Cvetic:2019gnh}
M.~Cveti\v{c}, J.~Halverson, L.~Lin, M.~Liu and J.~Tian,  {\em {Quadrillion
  $F$-Theory Compactifications with the Exact Chiral Spectrum of the Standard
  Model}}, Phys. Rev. Lett. {\bf 123} (2019), no.~10, 101601
[\href{http://www.arXiv.org/abs/1903.00009}{{\tt 1903.00009}}].
%%CITATION = ARXIV:1903.00009;%%.

\bibitem{Mayrhofer:2013ara}
C.~Mayrhofer, E.~Palti and T.~Weigand,  {\em {Hypercharge Flux in IIB and
  F-theory: Anomalies and Gauge Coupling Unification}}, JHEP {\bf 09} (2013)
  082 [\href{http://www.arXiv.org/abs/1303.3589}{{\tt 1303.3589}}].

\bibitem{Braun:2014pva}
A.~P. Braun, A.~Collinucci and R.~Valandro,  {\em {Hypercharge flux in F-theory
  and the stable Sen limit}}, JHEP {\bf 07} (2014) 121
  [\href{http://www.arXiv.org/abs/1402.4096}{{\tt 1402.4096}}].

\bibitem{Heckman:2010bq}
J.~J. Heckman,  {\em {Particle Physics Implications of F-theory}}, Ann. Rev.
  Nucl. Part. Sci. {\bf 60} (2010) 237--265
  [\href{http://www.arXiv.org/abs/1001.0577}{{\tt 1001.0577}}].

\bibitem{Schafer-Nameki:2015bva}
S.~Sch\"afer-Nameki,  {\em {F-theory: From Geometry to Phenomenology}}, Adv.
  Ser. Direct. High Energy Phys. {\bf 22} (2015) 245--275.

\bibitem{Weigand:2018rez}
T.~Weigand,  {\em {F-theory}}, PoS {\bf TASI2017} (2018) 016
  [\href{http://www.arXiv.org/abs/1806.01854}{{\tt 1806.01854}}].

\bibitem{BraunWatariGenerations}
A.~P. Braun and T.~Watari,  {\em {Distribution of the Number of Generations in
  Flux Compactifications}}, Phys. Rev. {\bf D90} (2014), no.~12, 121901
[\href{http://www.arXiv.org/abs/1408.6156}{{\tt 1408.6156}}].
%%CITATION = ARXIV:1408.6156;%%.

\bibitem{MorrisonTaylor4DClusters}
D.~R. Morrison and W.~Taylor,  {\em {Non-Higgsable clusters for 4D F-theory
  models}}, JHEP {\bf 05} (2015) 080
[\href{http://www.arXiv.org/abs/1412.6112}{{\tt 1412.6112}}].
%%CITATION = ARXIV:1412.6112;%%.

\bibitem{TianWangEString}
J.~Tian and Y.-N. Wang,  {\em {E-string and model building on a typical
  F-theory geometry}},
\href{http://www.arXiv.org/abs/[1811.02837]}{{\tt [1811.02837]}}.
%%CITATION = ARXIV:1811.02837;%%.

\bibitem{Denef:2004ze}
F.~Denef and M.~R. Douglas,  {\em {Distributions of flux vacua}}, JHEP {\bf 05}
  (2004) 072 [\href{http://www.arXiv.org/abs/hep-th/0404116}{{\tt
  hep-th/0404116}}].

\bibitem{DenefLesHouches}
F.~Denef,  {\em {Les Houches Lectures on Constructing String Vacua}}, Les
  Houches {\bf 87} (2008) 483--610
[\href{http://www.arXiv.org/abs/0803.1194}{{\tt 0803.1194}}].
%%CITATION = ARXIV:0803.1194;%%.

\bibitem{Bies:2014sra}
M.~Bies, C.~Mayrhofer, C.~Pehle and T.~Weigand,  {\em {Chow groups, Deligne
  cohomology and massless matter in F-theory}},
\href{http://www.arXiv.org/abs/1402.5144}{{\tt 1402.5144}}.
%%CITATION = ARXIV:1402.5144;%%.

\bibitem{Bies:2017fam}
M.~Bies, C.~Mayrhofer and T.~Weigand,  {\em {Gauge Backgrounds and Zero-Mode
  Counting in F-Theory}}, JHEP {\bf 11} (2017) 081
[\href{http://www.arXiv.org/abs/1706.04616}{{\tt 1706.04616}}].
%%CITATION = ARXIV:1706.04616;%%.

\bibitem{Bies:2021nje}
M.~Bies, M.~Cveti\v{c}, R.~Donagi, M.~Liu and M.~Ong,  {\em {Root bundles and
  towards exact matter spectra of F-theory MSSMs}}, JHEP {\bf 09} (2021) 076
  [\href{http://www.arXiv.org/abs/2102.10115}{{\tt 2102.10115}}].

\bibitem{Heckman:2008qa}
J.~J. Heckman and C.~Vafa,  {\em {Flavor Hierarchy From F-theory}}, Nucl. Phys.
  B {\bf 837} (2010) 137--151 [\href{http://www.arXiv.org/abs/0811.2417}{{\tt
  0811.2417}}].

\bibitem{Hayashi:2009ge}
H.~Hayashi, T.~Kawano, R.~Tatar and T.~Watari,  {\em {Codimension-3
  Singularities and Yukawa Couplings in F-theory}}, Nucl. Phys. B {\bf 823}
  (2009) 47--115 [\href{http://www.arXiv.org/abs/0901.4941}{{\tt 0901.4941}}].

\bibitem{Cecotti:2009zf}
S.~Cecotti, M.~C.~N. Cheng, J.~J. Heckman and C.~Vafa,  {\em {Yukawa Couplings
  in F-theory and Non-Commutative Geometry}},
  \href{http://www.arXiv.org/abs/0910.0477}{{\tt 0910.0477}}.

\bibitem{Cvetic:2019sgs}
M.~Cveti\v{c}, L.~Lin, M.~Liu, H.~Y. Zhang and G.~Zoccarato,  {\em {Yukawa
  Hierarchies in Global F-theory Models}}, JHEP {\bf 01} (2020) 037
  [\href{http://www.arXiv.org/abs/1906.10119}{{\tt 1906.10119}}].

\bibitem{Collinucci:2016hgh}
A.~Collinucci and I.~n. Garc\'\i{}a-Etxebarria,  {\em {E$_{6}$ Yukawa couplings
  in F-theory as D-brane instanton effects}}, JHEP {\bf 03} (2017) 155
  [\href{http://www.arXiv.org/abs/1612.06874}{{\tt 1612.06874}}].

\bibitem{Blumenhagen:2006xt}
R.~Blumenhagen, M.~Cvetic and T.~Weigand,  {\em {Spacetime instanton
  corrections in 4D string vacua: The Seesaw mechanism for D-Brane models}},
  Nucl. Phys. B {\bf 771} (2007) 113--142
  [\href{http://www.arXiv.org/abs/hep-th/0609191}{{\tt hep-th/0609191}}].

\bibitem{Ibanez:2006da}
L.~E. Ibanez and A.~M. Uranga,  {\em {Neutrino Majorana Masses from String
  Theory Instanton Effects}}, JHEP {\bf 03} (2007) 052
  [\href{http://www.arXiv.org/abs/hep-th/0609213}{{\tt hep-th/0609213}}].

\bibitem{Florea:2006si}
B.~Florea, S.~Kachru, J.~McGreevy and N.~Saulina,  {\em {Stringy Instantons and
  Quiver Gauge Theories}}, JHEP {\bf 05} (2007) 024
  [\href{http://www.arXiv.org/abs/hep-th/0610003}{{\tt hep-th/0610003}}].

\bibitem{Blumenhagen:2007zk}
R.~Blumenhagen, M.~Cvetic, D.~Lust, R.~Richter and T.~Weigand,  {\em
  {Non-perturbative Yukawa Couplings from String Instantons}}, Phys. Rev. Lett.
  {\bf 100} (2008) 061602 [\href{http://www.arXiv.org/abs/0707.1871}{{\tt
  0707.1871}}].

\bibitem{Blumenhagen:2009qh}
R.~Blumenhagen, M.~Cvetic, S.~Kachru and T.~Weigand,  {\em {D-Brane Instantons
  in Type II Orientifolds}}, Ann. Rev. Nucl. Part. Sci. {\bf 59} (2009)
  269--296 [\href{http://www.arXiv.org/abs/0902.3251}{{\tt 0902.3251}}].

\bibitem{Kakushadze:1997ub}
Z.~Kakushadze, G.~Shiu and S.~H.~H. Tye,  {\em {Couplings in asymmetric
  orbifolds and grand unified string models}}, Nucl. Phys. B {\bf 501} (1997)
  547--597 [\href{http://www.arXiv.org/abs/hep-th/9704113}{{\tt
  hep-th/9704113}}].

\bibitem{Braun:2008jp}
V.~Braun, T.~Brelidze, M.~R. Douglas and B.~A. Ovrut,  {\em {Eigenvalues and
  Eigenfunctions of the Scalar Laplace Operator on Calabi-Yau Manifolds}}, JHEP
  {\bf 07} (2008) 120 [\href{http://www.arXiv.org/abs/0805.3689}{{\tt
  0805.3689}}].

\bibitem{Cvetic:2003ch}
M.~Cveti{\v c} and I.~Papadimitriou,  {\em {Conformal field theory couplings
  for intersecting D-branes on orientifolds}}, Phys. Rev. D {\bf 68} (2003)
  046001 [\href{http://www.arXiv.org/abs/hep-th/0303083}{{\tt
  hep-th/0303083}}], [Erratum: Phys.Rev.D 70, 029903 (2004)].

\bibitem{Cvetic:2009yh}
M.~Cveti{\v c}, J.~Halverson and R.~Richter,  {\em {Realistic Yukawa structures
  from orientifold compactifications}}, JHEP {\bf 12} (2009) 063
  [\href{http://www.arXiv.org/abs/0905.3379}{{\tt 0905.3379}}].

\bibitem{Dienes:1996yh}
K.~R. Dienes and J.~March-Russell,  {\em {Realizing higher level gauge
  symmetries in string theory: New embeddings for string GUTs}}, Nucl. Phys. B
  {\bf 479} (1996) 113--172
  [\href{http://www.arXiv.org/abs/hep-th/9604112}{{\tt hep-th/9604112}}].

\bibitem{Klevers:2017aku}
D.~Klevers, D.~R. Morrison, N.~Raghuram and W.~Taylor,  {\em {Exotic matter on
  singular divisors in F-theory}}, JHEP {\bf 11} (2017) 124
  [\href{http://www.arXiv.org/abs/1706.08194}{{\tt 1706.08194}}].

\bibitem{Cvetic:2018xaq}
M.~Cveti\v{c}, J.~J. Heckman and L.~Lin,  {\em {Towards Exotic Matter and
  Discrete Non-Abelian Symmetries in F-theory}}, JHEP {\bf 11} (2018) 001
  [\href{http://www.arXiv.org/abs/1806.10594}{{\tt 1806.10594}}].

\bibitem{Esole:2020tby}
M.~Esole and M.~J. Kang,  {\em {Matter representations from geometry: under the
  spell of Dynkin}}, \href{http://www.arXiv.org/abs/2012.13401}{{\tt
  2012.13401}}.

\bibitem{Taylor:2018khc}
W.~Taylor and A.~P. Turner,  {\em {An infinite swampland of U(1) charge spectra
  in 6D supergravity theories}}, JHEP {\bf 06} (2018) 010
  [\href{http://www.arXiv.org/abs/1803.04447}{{\tt 1803.04447}}].

\bibitem{Raghuram:2018hjn}
N.~Raghuram and W.~Taylor,  {\em {Large U(1) charges in F-theory}}, JHEP {\bf
  10} (2018) 182 [\href{http://www.arXiv.org/abs/1809.01666}{{\tt
  1809.01666}}].

\bibitem{Cianci:2018vwv}
F.~M. Cianci, D.~K. Mayorga Pe\~na and R.~Valandro,  {\em {High U(1) charges in
  type IIB models and their F-theory lift}}, JHEP {\bf 04} (2019) 012
  [\href{http://www.arXiv.org/abs/1811.11777}{{\tt 1811.11777}}].

\bibitem{Collinucci:2019fnh}
A.~Collinucci, M.~Fazzi, D.~R. Morrison and R.~Valandro,  {\em {High electric
  charges in M-theory from quiver varieties}}, JHEP {\bf 11} (2019) 111
  [\href{http://www.arXiv.org/abs/1906.02202}{{\tt 1906.02202}}].

\bibitem{Raghuram:2021wvx}
N.~Raghuram and A.~P. Turner,  {\em {Orders of vanishing and U(1) charges in
  F-theory}}, JHEP {\bf 03} (2022) 051
  [\href{http://www.arXiv.org/abs/2110.10159}{{\tt 2110.10159}}].

\bibitem{Cvetic:2017epq}
M.~Cveti\v{c} and L.~Lin,  {\em {The Global Gauge Group Structure of F-theory
  Compactification with U(1)s}}, JHEP {\bf 01} (2018) 157
  [\href{http://www.arXiv.org/abs/1706.08521}{{\tt 1706.08521}}].

\bibitem{Taylor:2019ots}
W.~Taylor and A.~P. Turner,  {\em {Generic matter representations in 6D
  supergravity theories}}, JHEP {\bf 05} (2019) 081
  [\href{http://www.arXiv.org/abs/1901.02012}{{\tt 1901.02012}}].

\bibitem{Mayrhofer:2014opa}
C.~Mayrhofer, D.~R. Morrison, O.~Till and T.~Weigand,  {\em {Mordell-Weil
  Torsion and the Global Structure of Gauge Groups in F-theory}}, JHEP {\bf 10}
  (2014) 016 [\href{http://www.arXiv.org/abs/1405.3656}{{\tt 1405.3656}}].

\bibitem{Cvetic:2018bni}
M.~Cveti\v{c} and L.~Lin,  {\em {TASI Lectures on Abelian and Discrete
  Symmetries in F-theory}}, PoS {\bf TASI2017} (2018) 020
  [\href{http://www.arXiv.org/abs/1809.00012}{{\tt 1809.00012}}].

\bibitem{Banks:2010zn}
T.~Banks and N.~Seiberg,  {\em {Symmetries and Strings in Field Theory and
  Gravity}}, Phys. Rev. D {\bf 83} (2011) 084019
  [\href{http://www.arXiv.org/abs/1011.5120}{{\tt 1011.5120}}].

\bibitem{Harlow:2018tng}
D.~Harlow and H.~Ooguri,  {\em {Symmetries in quantum field theory and quantum
  gravity}}, Commun. Math. Phys. {\bf 383} (2021), no.~3, 1669--1804
  [\href{http://www.arXiv.org/abs/1810.05338}{{\tt 1810.05338}}].

\bibitem{Halverson:2018vbo}
J.~Halverson and P.~Langacker,  {\em {TASI Lectures on Remnants from the String
  Landscape}}, PoS {\bf TASI2017} (2018) 019
  [\href{http://www.arXiv.org/abs/1801.03503}{{\tt 1801.03503}}].

\bibitem{Svrcek:2006yi}
P.~Svrcek and E.~Witten,  {\em {Axions In String Theory}}, JHEP {\bf 06} (2006)
  051 [\href{http://www.arXiv.org/abs/hep-th/0605206}{{\tt hep-th/0605206}}].

\bibitem{Kreuzer:2000xy}
M.~Kreuzer and H.~Skarke,  {\em {Complete classification of reflexive polyhedra
  in four-dimensions}}, Adv. Theor. Math. Phys. {\bf 4} (2000) 1209--1230
  [\href{http://www.arXiv.org/abs/hep-th/0002240}{{\tt hep-th/0002240}}].

\bibitem{Taylor:2015xtz}
W.~Taylor and Y.-N. Wang,  {\em {The F-theory geometry with most flux vacua}},
  JHEP {\bf 12} (2015) 164 [\href{http://www.arXiv.org/abs/1511.03209}{{\tt
  1511.03209}}].

\bibitem{Taylor:2015ppa}
W.~Taylor and Y.-N. Wang,  {\em {A Monte Carlo exploration of threefold base
  geometries for 4d F-theory vacua}}, JHEP {\bf 01} (2016) 137
  [\href{http://www.arXiv.org/abs/1510.04978}{{\tt 1510.04978}}].

\bibitem{Halverson:2017ffz}
J.~Halverson, C.~Long and B.~Sung,  {\em {Algorithmic universality in F-theory
  compactifications}}, Phys. Rev. D {\bf 96} (2017), no.~12, 126006
  [\href{http://www.arXiv.org/abs/1706.02299}{{\tt 1706.02299}}].

\bibitem{Taylor:2017yqr}
W.~Taylor and Y.-N. Wang,  {\em {Scanning the skeleton of the 4D F-theory
  landscape}}, JHEP {\bf 01} (2018) 111
  [\href{http://www.arXiv.org/abs/1710.11235}{{\tt 1710.11235}}].

\bibitem{Arvanitaki:2009fg}
A.~Arvanitaki, S.~Dimopoulos, S.~Dubovsky, N.~Kaloper and J.~March-Russell,
  {\em {String Axiverse}}, Phys. Rev. D {\bf 81} (2010) 123530
  [\href{http://www.arXiv.org/abs/0905.4720}{{\tt 0905.4720}}].

\bibitem{Demirtas:2018akl}
M.~Demirtas, C.~Long, L.~McAllister and M.~Stillman,  {\em {The Kreuzer-Skarke
  Axiverse}}, JHEP {\bf 04} (2020) 138
  [\href{http://www.arXiv.org/abs/1808.01282}{{\tt 1808.01282}}].

\bibitem{McAllister:2008hb}
L.~McAllister, E.~Silverstein and A.~Westphal,  {\em {Gravity Waves and Linear
  Inflation from Axion Monodromy}}, Phys. Rev. D {\bf 82} (2010) 046003
  [\href{http://www.arXiv.org/abs/0808.0706}{{\tt 0808.0706}}].

\bibitem{Flauger:2009ab}
R.~Flauger, L.~McAllister, E.~Pajer, A.~Westphal and G.~Xu,  {\em {Oscillations
  in the CMB from Axion Monodromy Inflation}}, JCAP {\bf 06} (2010) 009
  [\href{http://www.arXiv.org/abs/0907.2916}{{\tt 0907.2916}}].

\bibitem{Marchesano:2014mla}
F.~Marchesano, G.~Shiu and A.~M. Uranga,  {\em {F-term Axion Monodromy
  Inflation}}, JHEP {\bf 09} (2014) 184
  [\href{http://www.arXiv.org/abs/1404.3040}{{\tt 1404.3040}}].

\bibitem{Hebecker:2014eua}
A.~Hebecker, S.~C. Kraus and L.~T. Witkowski,  {\em {D7-Brane Chaotic
  Inflation}}, Phys. Lett. B {\bf 737} (2014) 16--22
  [\href{http://www.arXiv.org/abs/1404.3711}{{\tt 1404.3711}}].

\bibitem{McAllister:2014mpa}
L.~McAllister, E.~Silverstein, A.~Westphal and T.~Wrase,  {\em {The Powers of
  Monodromy}}, JHEP {\bf 09} (2014) 123
  [\href{http://www.arXiv.org/abs/1405.3652}{{\tt 1405.3652}}].

\bibitem{Hebecker:2014kva}
A.~Hebecker, P.~Mangat, F.~Rompineve and L.~T. Witkowski,  {\em {Tuning and
  Backreaction in F-term Axion Monodromy Inflation}}, Nucl. Phys. B {\bf 894}
  (2015) 456--495 [\href{http://www.arXiv.org/abs/1411.2032}{{\tt 1411.2032}}].

\bibitem{Blumenhagen:2014gta}
R.~Blumenhagen and E.~Plauschinn,  {\em {Towards Universal Axion Inflation and
  Reheating in String Theory}}, Phys. Lett. B {\bf 736} (2014) 482--487
  [\href{http://www.arXiv.org/abs/1404.3542}{{\tt 1404.3542}}].

\bibitem{Halverson:2019kna}
J.~Halverson, C.~Long, B.~Nelson and G.~Salinas,  {\em {Axion reheating in the
  string landscape}}, Phys. Rev. D {\bf 99} (2019), no.~8, 086014
  [\href{http://www.arXiv.org/abs/1903.04495}{{\tt 1903.04495}}].

\bibitem{Ibanez:2015fcv}
L.~E. Ibanez, M.~Montero, A.~Uranga and I.~Valenzuela,  {\em {Relaxion
  Monodromy and the Weak Gravity Conjecture}}, JHEP {\bf 04} (2016) 020
  [\href{http://www.arXiv.org/abs/1512.00025}{{\tt 1512.00025}}].

\bibitem{Brown:2016nqt}
J.~Brown, W.~Cottrell, G.~Shiu and P.~Soler,  {\em {Tunneling in Axion
  Monodromy}}, JHEP {\bf 10} (2016) 025
  [\href{http://www.arXiv.org/abs/1607.00037}{{\tt 1607.00037}}].

\bibitem{McAllister:2016vzi}
L.~McAllister, P.~Schwaller, G.~Servant, J.~Stout and A.~Westphal,  {\em
  {Runaway Relaxion Monodromy}}, JHEP {\bf 02} (2018) 124
  [\href{http://www.arXiv.org/abs/1610.05320}{{\tt 1610.05320}}].

\bibitem{Hui:2016ltb}
L.~Hui, J.~P. Ostriker, S.~Tremaine and E.~Witten,  {\em {Ultralight scalars as
  cosmological dark matter}}, Phys. Rev. D {\bf 95} (2017), no.~4, 043541
  [\href{http://www.arXiv.org/abs/1610.08297}{{\tt 1610.08297}}].

\bibitem{Halverson:2017deq}
J.~Halverson, C.~Long and P.~Nath,  {\em {Ultralight axion in supersymmetry and
  strings and cosmology at small scales}}, Phys. Rev. D {\bf 96} (2017), no.~5,
  056025 [\href{http://www.arXiv.org/abs/1703.07779}{{\tt 1703.07779}}].

\bibitem{Cicoli:2021gss}
M.~Cicoli, V.~Guidetti, N.~Righi and A.~Westphal,  {\em {Fuzzy Dark Matter
  Candidates from String Theory}},
  \href{http://www.arXiv.org/abs/2110.02964}{{\tt 2110.02964}}.

\bibitem{Mehta:2021pwf}
V.~M. Mehta, M.~Demirtas, C.~Long, D.~J.~E. Marsh, L.~McAllister and M.~J.
  Stott,  {\em {Superradiance in string theory}}, JCAP {\bf 07} (2021) 033
  [\href{http://www.arXiv.org/abs/2103.06812}{{\tt 2103.06812}}].

\bibitem{Kitajima:2018zco}
N.~Kitajima, J.~Soda and Y.~Urakawa,  {\em {Gravitational wave forest from
  string axiverse}}, JCAP {\bf 10} (2018) 008
  [\href{http://www.arXiv.org/abs/1807.07037}{{\tt 1807.07037}}].

\bibitem{Barnaby:2012xt}
N.~Barnaby, J.~Moxon, R.~Namba, M.~Peloso, G.~Shiu and P.~Zhou,  {\em {Gravity
  waves and non-Gaussian features from particle production in a sector
  gravitationally coupled to the inflaton}}, Phys. Rev. D {\bf 86} (2012)
  103508 [\href{http://www.arXiv.org/abs/1206.6117}{{\tt 1206.6117}}].

\bibitem{Mukohyama:2014gba}
S.~Mukohyama, R.~Namba, M.~Peloso and G.~Shiu,  {\em {Blue Tensor Spectrum from
  Particle Production during Inflation}}, JCAP {\bf 08} (2014) 036
  [\href{http://www.arXiv.org/abs/1405.0346}{{\tt 1405.0346}}].

\bibitem{Liu:2017hua}
J.~Liu, Z.-K. Guo, R.-G. Cai and G.~Shiu,  {\em {Gravitational Waves from
  Oscillons with Cuspy Potentials}}, Phys. Rev. Lett. {\bf 120} (2018), no.~3,
  031301 [\href{http://www.arXiv.org/abs/1707.09841}{{\tt 1707.09841}}].

\bibitem{Liu:2018rrt}
J.~Liu, Z.-K. Guo, R.-G. Cai and G.~Shiu,  {\em {Gravitational wave production
  after inflation with cuspy potentials}}, Phys. Rev. D {\bf 99} (2018),
  no.~10, 103506 [\href{http://www.arXiv.org/abs/1812.09235}{{\tt
  1812.09235}}].

\bibitem{Halverson:2019cmy}
J.~Halverson, C.~Long, B.~Nelson and G.~Salinas,  {\em {Towards string theory
  expectations for photon couplings to axionlike particles}}, Phys. Rev. D {\bf
  100} (2019), no.~10, 106010 [\href{http://www.arXiv.org/abs/1909.05257}{{\tt
  1909.05257}}].

\bibitem{Demirtas:2021gsq}
M.~Demirtas, N.~Gendler, C.~Long, L.~McAllister and J.~Moritz,  {\em {PQ
  Axiverse}}, \href{http://www.arXiv.org/abs/2112.04503}{{\tt 2112.04503}}.

\bibitem{Faraggi:2000pv}
A.~E. Faraggi and M.~Pospelov,  {\em {Self-interacting dark matter from the
  hidden heterotic string sector}}, Astropart. Phys. {\bf 16} (2002) 451--461
  [\href{http://www.arXiv.org/abs/hep-ph/0008223}{{\tt hep-ph/0008223}}].

\bibitem{Friedman:1997yq}
R.~Friedman, J.~Morgan and E.~Witten,  {\em {Vector bundles and F theory}},
  Commun. Math. Phys. {\bf 187} (1997) 679--743
  [\href{http://www.arXiv.org/abs/hep-th/9701162}{{\tt hep-th/9701162}}].

\bibitem{Braun:2017uku}
A.~P. Braun and S.~Sch\"afer-Nameki,  {\em {Compact, Singular $G_2$-Holonomy
  Manifolds and M/Heterotic/F-Theory Duality}}, JHEP {\bf 04} (2018) 126
  [\href{http://www.arXiv.org/abs/1708.07215}{{\tt 1708.07215}}].

\bibitem{Halverson:2016nfq}
J.~Halverson, B.~D. Nelson and F.~Ruehle,  {\em {String Theory and the Dark
  Glueball Problem}}, Phys. Rev. D {\bf 95} (2017), no.~4, 043527
  [\href{http://www.arXiv.org/abs/1609.02151}{{\tt 1609.02151}}].

\bibitem{Dienes:2016vei}
K.~R. Dienes, F.~Huang, S.~Su and B.~Thomas,  {\em {Dynamical Dark Matter from
  Strongly-Coupled Dark Sectors}}, Phys. Rev. D {\bf 95} (2017), no.~4, 043526
  [\href{http://www.arXiv.org/abs/1610.04112}{{\tt 1610.04112}}].

\bibitem{Acharya:2017szw}
B.~S. Acharya, M.~Fairbairn and E.~Hardy,  {\em {Glueball dark matter in
  non-standard cosmologies}}, JHEP {\bf 07} (2017) 100
  [\href{http://www.arXiv.org/abs/1704.01804}{{\tt 1704.01804}}].

\bibitem{Feng:2014cla}
W.-Z. Feng, G.~Shiu, P.~Soler and F.~Ye,  {\em {Building a St\"uckelberg
  portal}}, JHEP {\bf 05} (2014) 065
  [\href{http://www.arXiv.org/abs/1401.5890}{{\tt 1401.5890}}].

\bibitem{Feng:2014eja}
W.-Z. Feng, G.~Shiu, P.~Soler and F.~Ye,  {\em {Probing Hidden Sectors with
  St\"uckelberg U(1) Gauge Fields}}, Phys. Rev. Lett. {\bf 113} (2014) 061802
  [\href{http://www.arXiv.org/abs/1401.5880}{{\tt 1401.5880}}].

\bibitem{Cirelli:2005uq}
M.~Cirelli, N.~Fornengo and A.~Strumia,  {\em {Minimal dark matter}}, Nucl.
  Phys. B {\bf 753} (2006) 178--194
  [\href{http://www.arXiv.org/abs/hep-ph/0512090}{{\tt hep-ph/0512090}}].

\bibitem{Halverson:2014nwa}
J.~Halverson, N.~Orlofsky and A.~Pierce,  {\em {Vectorlike Leptons as the Tip
  of the Dark Matter Iceberg}}, Phys. Rev. D {\bf 90} (2014), no.~1, 015002
  [\href{http://www.arXiv.org/abs/1403.1592}{{\tt 1403.1592}}].

\bibitem{LHCb:2021trn}
{LHCb} Collaboration, R.~Aaij {\em et al.},  {\em {Test of lepton universality
  in beauty-quark decays}}, Nature Phys. {\bf 18} (2022), no.~3, 277--282
  [\href{http://www.arXiv.org/abs/2103.11769}{{\tt 2103.11769}}].

\bibitem{Anastasopoulos:2006da}
P.~Anastasopoulos, T.~P.~T. Dijkstra, E.~Kiritsis and A.~N. Schellekens,  {\em
  {Orientifolds, hypercharge embeddings and the Standard Model}}, Nucl. Phys. B
  {\bf 759} (2006) 83--146 [\href{http://www.arXiv.org/abs/hep-th/0605226}{{\tt
  hep-th/0605226}}].

\bibitem{Cvetic:2011iq}
M.~Cveti\v{c}, J.~Halverson and P.~Langacker,  {\em {Implications of String
  Constraints for Exotic Matter and Z' s Beyond the Standard Model}}, JHEP {\bf
  11} (2011) 058 [\href{http://www.arXiv.org/abs/1108.5187}{{\tt 1108.5187}}].

\bibitem{Halverson:2013ska}
J.~Halverson,  {\em {Anomaly Nucleation Constrains SU(2) Gauge Theories}},
  Phys. Rev. Lett. {\bf 111} (2013), no.~26, 261601
  [\href{http://www.arXiv.org/abs/1310.1091}{{\tt 1310.1091}}].

\bibitem{Kakushadze:1997mc}
Z.~Kakushadze, G.~Shiu, S.~H.~H. Tye and Y.~Vtorov-Karevsky,  {\em {A Review of
  three family grand unified string models}}, Int. J. Mod. Phys. A {\bf 13}
  (1998) 2551--2598 [\href{http://www.arXiv.org/abs/hep-th/9710149}{{\tt
  hep-th/9710149}}].

\bibitem{Cvetic:2002qa}
M.~Cvetic, P.~Langacker and G.~Shiu,  {\em {Phenomenology of a three family
  standard like string model}}, Phys. Rev. D {\bf 66} (2002) 066004
  [\href{http://www.arXiv.org/abs/hep-ph/0205252}{{\tt hep-ph/0205252}}].

\bibitem{MorrisonTaylorClusters}
D.~R. Morrison and W.~Taylor,  {\em {Classifying bases for 6D F-theory
  models}}, Central Eur. J. Phys. {\bf 10} (2012) 1072--1088
[\href{http://www.arXiv.org/abs/1201.1943}{{\tt 1201.1943}}].
%%CITATION = ARXIV:1201.1943;%%.

\bibitem{Acharya:2019mcu}
B.~S. Acharya,  {\em {Supersymmetry, Ricci Flat Manifolds and the String
  Landscape}}, JHEP {\bf 08} (2020) 128
  [\href{http://www.arXiv.org/abs/1906.06886}{{\tt 1906.06886}}].

\bibitem{Witten:1981gj}
E.~Witten,  {\em {Instability of the Kaluza-Klein Vacuum}}, Nucl. Phys. B {\bf
  195} (1982) 481--492.

\bibitem{GarciaEtxebarria:2020xsr}
I.~n. Garc\'\i{}a~Etxebarria, M.~Montero, K.~Sousa and I.~Valenzuela,  {\em
  {Nothing is certain in string compactifications}}, JHEP {\bf 12} (2020) 032
  [\href{http://www.arXiv.org/abs/2005.06494}{{\tt 2005.06494}}].

\bibitem{Acharya:2020hsc}
B.~S. Acharya, G.~Aldazabal, E.~Andr\'es, A.~Font, K.~Narain and I.~G. Zadeh,
  {\em {Stringy Tachyonic Instabilities of Non-Supersymmetric Ricci Flat
  Backgrounds}}, JHEP {\bf 04} (2021) 026
  [\href{http://www.arXiv.org/abs/2010.02933}{{\tt 2010.02933}}].

\bibitem{DeLuca:2021pej}
G.~B. De~Luca, E.~Silverstein and G.~Torroba,  {\em {Hyperbolic
  compactification of M-theory and de Sitter quantum gravity}}, SciPost Phys.
  {\bf 12} (2022), no.~3, 083 [\href{http://www.arXiv.org/abs/2104.13380}{{\tt
  2104.13380}}].

\bibitem{Blumenhagen:2007sm}
R.~Blumenhagen, S.~Moster and E.~Plauschinn,  {\em {Moduli Stabilisation versus
  Chirality for MSSM like Type IIB Orientifolds}}, JHEP {\bf 01} (2008) 058
  [\href{http://www.arXiv.org/abs/0711.3389}{{\tt 0711.3389}}].

\bibitem{Blumenhagen:2009gk}
R.~Blumenhagen, J.~P. Conlon, S.~Krippendorf, S.~Moster and F.~Quevedo,  {\em
  {SUSY Breaking in Local String/F-Theory Models}}, JHEP {\bf 09} (2009) 007
  [\href{http://www.arXiv.org/abs/0906.3297}{{\tt 0906.3297}}].

\bibitem{Grimm:2011tb}
T.~W. Grimm, M.~Kerstan, E.~Palti and T.~Weigand,  {\em {Massive Abelian Gauge
  Symmetries and Fluxes in F-theory}}, JHEP {\bf 12} (2011) 004
  [\href{http://www.arXiv.org/abs/1107.3842}{{\tt 1107.3842}}].

\bibitem{Cicoli:2012vw}
M.~Cicoli, S.~Krippendorf, C.~Mayrhofer, F.~Quevedo and R.~Valandro,  {\em
  {D-Branes at del Pezzo Singularities: Global Embedding and Moduli
  Stabilisation}}, JHEP {\bf 09} (2012) 019
  [\href{http://www.arXiv.org/abs/1206.5237}{{\tt 1206.5237}}].

\bibitem{Carta:2019rhx}
F.~Carta, J.~Moritz and A.~Westphal,  {\em {Gaugino condensation and small
  uplifts in KKLT}}, JHEP {\bf 08} (2019) 141
  [\href{http://www.arXiv.org/abs/1902.01412}{{\tt 1902.01412}}].

\bibitem{Bena:2020xrh}
I.~Bena, J.~Bl\r{a}b\"ack, M.~Gra\~na and S.~L\"ust,  {\em {The tadpole
  problem}}, JHEP {\bf 11} (2021) 223
  [\href{http://www.arXiv.org/abs/2010.10519}{{\tt 2010.10519}}].

\bibitem{Bena:2021wyr}
I.~Bena, J.~Bl\r{a}b\"ack, M.~Gra\~na and S.~L\"ust,  {\em {Algorithmically
  Solving the Tadpole Problem}}, Adv. Appl. Clifford Algebras {\bf 32} (2022),
  no.~1, 7 [\href{http://www.arXiv.org/abs/2103.03250}{{\tt 2103.03250}}].

\bibitem{Gao:2022fdi}
X.~Gao, A.~Hebecker, S.~Schreyer and G.~Venken,  {\em {The LVS Parametric
  Tadpole Constraint}}, \href{http://www.arXiv.org/abs/2202.04087}{{\tt
  2202.04087}}.

\bibitem{Douglas:2006es}
M.~R. Douglas and S.~Kachru,  {\em {Flux compactification}}, Rev. Mod. Phys.
  {\bf 79} (2007) 733--796 [\href{http://www.arXiv.org/abs/hep-th/0610102}{{\tt
  hep-th/0610102}}].

\bibitem{Kachru:2003aw}
S.~Kachru, R.~Kallosh, A.~D. Linde and S.~P. Trivedi,  {\em {De Sitter vacua in
  string theory}}, Phys. Rev. D {\bf 68} (2003) 046005
  [\href{http://www.arXiv.org/abs/hep-th/0301240}{{\tt hep-th/0301240}}].

\bibitem{Balasubramanian:2005zx}
V.~Balasubramanian, P.~Berglund, J.~P. Conlon and F.~Quevedo,  {\em
  {Systematics of moduli stabilisation in Calabi-Yau flux compactifications}},
  JHEP {\bf 03} (2005) 007 [\href{http://www.arXiv.org/abs/hep-th/0502058}{{\tt
  hep-th/0502058}}].

\bibitem{Maldacena:2000mw}
J.~M. Maldacena and C.~Nunez,  {\em {Supergravity description of field theories
  on curved manifolds and a no go theorem}}, Int. J. Mod. Phys. A {\bf 16}
  (2001) 822--855 [\href{http://www.arXiv.org/abs/hep-th/0007018}{{\tt
  hep-th/0007018}}].

\bibitem{Hertzberg:2007wc}
M.~P. Hertzberg, S.~Kachru, W.~Taylor and M.~Tegmark,  {\em {Inflationary
  Constraints on Type IIA String Theory}}, JHEP {\bf 12} (2007) 095
  [\href{http://www.arXiv.org/abs/0711.2512}{{\tt 0711.2512}}].

\bibitem{Danielsson:2009ff}
U.~H. Danielsson, S.~S. Haque, G.~Shiu and T.~Van~Riet,  {\em {Towards
  Classical de Sitter Solutions in String Theory}}, JHEP {\bf 09} (2009) 114
  [\href{http://www.arXiv.org/abs/0907.2041}{{\tt 0907.2041}}].

\bibitem{Wrase:2010ew}
T.~Wrase and M.~Zagermann,  {\em {On Classical de Sitter Vacua in String
  Theory}}, Fortsch. Phys. {\bf 58} (2010) 906--910
  [\href{http://www.arXiv.org/abs/1003.0029}{{\tt 1003.0029}}].

\bibitem{Danielsson:2011au}
U.~H. Danielsson, S.~S. Haque, P.~Koerber, G.~Shiu, T.~Van~Riet and T.~Wrase,
  {\em {De Sitter hunting in a classical landscape}}, Fortsch. Phys. {\bf 59}
  (2011) 897--933 [\href{http://www.arXiv.org/abs/1103.4858}{{\tt 1103.4858}}].

\bibitem{Shiu:2011zt}
G.~Shiu and Y.~Sumitomo,  {\em {Stability Constraints on Classical de Sitter
  Vacua}}, JHEP {\bf 09} (2011) 052
  [\href{http://www.arXiv.org/abs/1107.2925}{{\tt 1107.2925}}].

\bibitem{Danielsson:2012et}
U.~H. Danielsson, G.~Shiu, T.~Van~Riet and T.~Wrase,  {\em {A note on obstinate
  tachyons in classical dS solutions}}, JHEP {\bf 03} (2013) 138
  [\href{http://www.arXiv.org/abs/1212.5178}{{\tt 1212.5178}}].

\bibitem{Junghans:2016uvg}
D.~Junghans,  {\em {Tachyons in Classical de Sitter Vacua}}, JHEP {\bf 06}
  (2016) 132 [\href{http://www.arXiv.org/abs/1603.08939}{{\tt 1603.08939}}].

\bibitem{Junghans:2016abx}
D.~Junghans and M.~Zagermann,  {\em {A Universal Tachyon in Nearly No-scale de
  Sitter Compactifications}}, JHEP {\bf 07} (2018) 078
  [\href{http://www.arXiv.org/abs/1612.06847}{{\tt 1612.06847}}].

\bibitem{Andriot:2018wzk}
D.~Andriot,  {\em {On the de Sitter swampland criterion}}, Phys. Lett. B {\bf
  785} (2018) 570--573 [\href{http://www.arXiv.org/abs/1806.10999}{{\tt
  1806.10999}}].

\bibitem{Andriot:2018ept}
D.~Andriot,  {\em {New constraints on classical de Sitter: flirting with the
  swampland}}, Fortsch. Phys. {\bf 67} (2019), no.~1-2, 1800103
  [\href{http://www.arXiv.org/abs/1807.09698}{{\tt 1807.09698}}].

\bibitem{Garg:2018reu}
S.~K. Garg and C.~Krishnan,  {\em {Bounds on Slow Roll and the de Sitter
  Swampland}}, JHEP {\bf 11} (2019) 075
  [\href{http://www.arXiv.org/abs/1807.05193}{{\tt 1807.05193}}].

\bibitem{Danielsson:2018ztv}
U.~H. Danielsson and T.~Van~Riet,  {\em {What if string theory has no de Sitter
  vacua?}}, Int. J. Mod. Phys. D {\bf 27} (2018), no.~12, 1830007
  [\href{http://www.arXiv.org/abs/1804.01120}{{\tt 1804.01120}}].

\bibitem{Obied:2018sgi}
G.~Obied, H.~Ooguri, L.~Spodyneiko and C.~Vafa,  {\em {De Sitter Space and the
  Swampland}}, \href{http://www.arXiv.org/abs/1806.08362}{{\tt 1806.08362}}.

\bibitem{Ooguri:2018wrx}
H.~Ooguri, E.~Palti, G.~Shiu and C.~Vafa,  {\em {Distance and de Sitter
  Conjectures on the Swampland}}, Phys. Lett. B {\bf 788} (2019) 180--184
  [\href{http://www.arXiv.org/abs/1810.05506}{{\tt 1810.05506}}].

\bibitem{Dine:1985he}
M.~Dine and N.~Seiberg,  {\em {Is the Superstring Weakly Coupled?}}, Phys.
  Lett. B {\bf 162} (1985) 299--302.

\bibitem{Freivogel:2016qwc}
B.~Freivogel and M.~Kleban,  {\em {Vacua Morghulis}},
  \href{http://www.arXiv.org/abs/1610.04564}{{\tt 1610.04564}}.

\bibitem{Sethi:2017phn}
S.~Sethi,  {\em {Supersymmetry Breaking by Fluxes}}, JHEP {\bf 10} (2018) 022
  [\href{http://www.arXiv.org/abs/1709.03554}{{\tt 1709.03554}}].

\bibitem{Demirtas:2021nlu}
M.~Demirtas, M.~Kim, L.~McAllister, J.~Moritz and A.~Rios-Tascon,  {\em {Small
  cosmological constants in string theory}}, JHEP {\bf 12} (2021) 136
  [\href{http://www.arXiv.org/abs/2107.09064}{{\tt 2107.09064}}].

\bibitem{Cole:2019enn}
A.~Cole, A.~Schachner and G.~Shiu,  {\em {Searching the Landscape of Flux Vacua
  with Genetic Algorithms}}, JHEP {\bf 11} (2019) 045
  [\href{http://www.arXiv.org/abs/1907.10072}{{\tt 1907.10072}}].

\bibitem{Gao:2020xqh}
X.~Gao, A.~Hebecker and D.~Junghans,  {\em {Control issues of KKLT}}, Fortsch.
  Phys. {\bf 68} (2020) 2000089
  [\href{http://www.arXiv.org/abs/2009.03914}{{\tt 2009.03914}}].

\bibitem{Carta:2021lqg}
F.~Carta and J.~Moritz,  {\em {Resolving spacetime singularities in flux
  compactifications \& KKLT}}, JHEP {\bf 08} (2021) 093
  [\href{http://www.arXiv.org/abs/2101.05281}{{\tt 2101.05281}}].

\bibitem{Hamada:2018qef}
Y.~Hamada, A.~Hebecker, G.~Shiu and P.~Soler,  {\em {On brane gaugino
  condensates in 10d}}, JHEP {\bf 04} (2019) 008
  [\href{http://www.arXiv.org/abs/1812.06097}{{\tt 1812.06097}}].

\bibitem{Hamada:2019ack}
Y.~Hamada, A.~Hebecker, G.~Shiu and P.~Soler,  {\em {Understanding KKLT from a
  10d perspective}}, JHEP {\bf 06} (2019) 019
  [\href{http://www.arXiv.org/abs/1902.01410}{{\tt 1902.01410}}].

\bibitem{Kachru:2019dvo}
S.~Kachru, M.~Kim, L.~Mcallister and M.~Zimet,  {\em {de Sitter vacua from ten
  dimensions}}, JHEP {\bf 12} (2021) 111
  [\href{http://www.arXiv.org/abs/1908.04788}{{\tt 1908.04788}}].

\bibitem{Hamada:2021ryq}
Y.~Hamada, A.~Hebecker, G.~Shiu and P.~Soler,  {\em {Completing the D7-brane
  local gaugino action}}, JHEP {\bf 11} (2021) 033
  [\href{http://www.arXiv.org/abs/2105.11467}{{\tt 2105.11467}}].

\bibitem{Bena:2019mte}
I.~Bena, M.~Gra\~na, N.~Kovensky and A.~Retolaza,  {\em {K\"ahler moduli
  stabilization from ten dimensions}}, JHEP {\bf 10} (2019) 200
  [\href{http://www.arXiv.org/abs/1908.01785}{{\tt 1908.01785}}].

\bibitem{Cvetic:2020fkd}
M.~Cveti{\v c}, J.~Halverson, L.~Lin and C.~Long,  {\em {Constraints on
  Standard Model Constructions in F-theory}}, Phys. Rev. D {\bf 102} (2020),
  no.~2, 026012 [\href{http://www.arXiv.org/abs/2004.00630}{{\tt 2004.00630}}].

\bibitem{tHooft:1979rat}
G.~'t~Hooft,  {\em {Naturalness, chiral symmetry, and spontaneous chiral
  symmetry breaking}}, NATO Sci. Ser. B {\bf 59} (1980) 135--157.

\bibitem{Banks:1988yz}
T.~Banks and L.~J. Dixon,  {\em {Constraints on String Vacua with Space-Time
  Supersymmetry}}, Nucl. Phys. B {\bf 307} (1988) 93--108.

\bibitem{Minwalla:1999px}
S.~Minwalla, M.~Van~Raamsdonk and N.~Seiberg,  {\em {Noncommutative
  perturbative dynamics}}, JHEP {\bf 02} (2000) 020
  [\href{http://www.arXiv.org/abs/hep-th/9912072}{{\tt hep-th/9912072}}].

\bibitem{Abel:2021tyt}
S.~Abel and K.~R. Dienes,  {\em {Calculating the Higgs mass in string theory}},
  Phys. Rev. D {\bf 104} (2021), no.~12, 126032
  [\href{http://www.arXiv.org/abs/2106.04622}{{\tt 2106.04622}}].

\bibitem{Douglas:2004zg}
M.~R. Douglas,  {\em {Basic results in vacuum statistics}}, Comptes Rendus
  Physique {\bf 5} (2004) 965--977
  [\href{http://www.arXiv.org/abs/hep-th/0409207}{{\tt hep-th/0409207}}].

\bibitem{Susskind:2004uv}
L.~Susskind,  {\em {Supersymmetry breaking in the anthropic landscape}}, in
  {\em {From Fields to Strings: Circumnavigating Theoretical Physics: A
  Conference in Tribute to Ian Kogan}}, pp.~1745--1749.
\newblock 5, 2004.
\newblock \href{http://www.arXiv.org/abs/hep-th/0405189}{{\tt hep-th/0405189}}.

\bibitem{Douglas:2004qg}
M.~R. Douglas,  {\em {Statistical analysis of the supersymmetry breaking
  scale}}, \href{http://www.arXiv.org/abs/hep-th/0405279}{{\tt
  hep-th/0405279}}.

\bibitem{Dine:2004is}
M.~Dine, E.~Gorbatov and S.~D. Thomas,  {\em {Low energy supersymmetry from the
  landscape}}, JHEP {\bf 08} (2008) 098
  [\href{http://www.arXiv.org/abs/hep-th/0407043}{{\tt hep-th/0407043}}].

\bibitem{Vafa:2005ui}
C.~Vafa,  {\em {The String landscape and the swampland}},
  \href{http://www.arXiv.org/abs/hep-th/0509212}{{\tt hep-th/0509212}}.

\bibitem{Brennan:2017rbf}
T.~D. Brennan, F.~Carta and C.~Vafa,  {\em {The String Landscape, the
  Swampland, and the Missing Corner}}, PoS {\bf TASI2017} (2017) 015
  [\href{http://www.arXiv.org/abs/1711.00864}{{\tt 1711.00864}}].

\bibitem{Palti:2019pca}
E.~Palti,  {\em {The Swampland: Introduction and Review}}, Fortsch. Phys. {\bf
  67} (2019), no.~6, 1900037 [\href{http://www.arXiv.org/abs/1903.06239}{{\tt
  1903.06239}}].

\bibitem{vanBeest:2021lhn}
M.~van Beest, J.~Calder\'on-Infante, D.~Mirfendereski and I.~Valenzuela,  {\em
  {Lectures on the Swampland Program in String Compactifications}},
  \href{http://www.arXiv.org/abs/2102.01111}{{\tt 2102.01111}}.

\bibitem{Grana:2021zvf}
M.~Gra\~na and A.~Herr\'aez,  {\em {The Swampland Conjectures: A Bridge from
  Quantum Gravity to Particle Physics}}, Universe {\bf 7} (2021), no.~8, 273
  [\href{http://www.arXiv.org/abs/2107.00087}{{\tt 2107.00087}}].

\bibitem{Harlow:2022gzl}
D.~Harlow, B.~Heidenreich, M.~Reece and T.~Rudelius,  {\em {The Weak Gravity
  Conjecture: A Review}}, \href{http://www.arXiv.org/abs/2201.08380}{{\tt
  2201.08380}}.

\bibitem{Shiu:2013wxa}
G.~Shiu, P.~Soler and F.~Ye,  {\em {Milli-Charged Dark Matter in Quantum
  Gravity and String Theory}}, Phys. Rev. Lett. {\bf 110} (2013), no.~24,
  241304 [\href{http://www.arXiv.org/abs/1302.5471}{{\tt 1302.5471}}].

\bibitem{Arkani-Hamed:2006emk}
N.~Arkani-Hamed, L.~Motl, A.~Nicolis and C.~Vafa,  {\em {The String landscape,
  black holes and gravity as the weakest force}}, JHEP {\bf 06} (2007) 060
  [\href{http://www.arXiv.org/abs/hep-th/0601001}{{\tt hep-th/0601001}}].

\bibitem{Brown:2015iha}
J.~Brown, W.~Cottrell, G.~Shiu and P.~Soler,  {\em {Fencing in the Swampland:
  Quantum Gravity Constraints on Large Field Inflation}}, JHEP {\bf 10} (2015)
  023 [\href{http://www.arXiv.org/abs/1503.04783}{{\tt 1503.04783}}].

\bibitem{Brown:2015lia}
J.~Brown, W.~Cottrell, G.~Shiu and P.~Soler,  {\em {On Axionic Field Ranges,
  Loopholes and the Weak Gravity Conjecture}}, JHEP {\bf 04} (2016) 017
  [\href{http://www.arXiv.org/abs/1504.00659}{{\tt 1504.00659}}].

\bibitem{Montero:2015ofa}
M.~Montero, A.~M. Uranga and I.~Valenzuela,  {\em {Transplanckian axions!?}},
  JHEP {\bf 08} (2015) 032 [\href{http://www.arXiv.org/abs/1503.03886}{{\tt
  1503.03886}}].

\bibitem{Heidenreich:2015wga}
B.~Heidenreich, M.~Reece and T.~Rudelius,  {\em {Weak Gravity Strongly
  Constrains Large-Field Axion Inflation}}, JHEP {\bf 12} (2015) 108
  [\href{http://www.arXiv.org/abs/1506.03447}{{\tt 1506.03447}}].

\bibitem{Hebecker:2018ofv}
A.~Hebecker, T.~Mikhail and P.~Soler,  {\em {Euclidean wormholes, baby
  universes, and their impact on particle physics and cosmology}}, Front.
  Astron. Space Sci. {\bf 5} (2018) 35
  [\href{http://www.arXiv.org/abs/1807.00824}{{\tt 1807.00824}}].

\bibitem{Reece:2018zvv}
M.~Reece,  {\em {Photon Masses in the Landscape and the Swampland}}, JHEP {\bf
  07} (2019) 181 [\href{http://www.arXiv.org/abs/1808.09966}{{\tt
  1808.09966}}].

\bibitem{Ibanez:2017kvh}
L.~E. Ibanez, V.~Martin-Lozano and I.~Valenzuela,  {\em {Constraining Neutrino
  Masses, the Cosmological Constant and BSM Physics from the Weak Gravity
  Conjecture}}, JHEP {\bf 11} (2017) 066
  [\href{http://www.arXiv.org/abs/1706.05392}{{\tt 1706.05392}}].

\bibitem{Hamada:2017yji}
Y.~Hamada and G.~Shiu,  {\em {Weak Gravity Conjecture, Multiple Point Principle
  and the Standard Model Landscape}}, JHEP {\bf 11} (2017) 043
  [\href{http://www.arXiv.org/abs/1707.06326}{{\tt 1707.06326}}].

\bibitem{Lust:2017wrl}
D.~Lust and E.~Palti,  {\em {Scalar Fields, Hierarchical UV/IR Mixing and The
  Weak Gravity Conjecture}}, JHEP {\bf 02} (2018) 040
  [\href{http://www.arXiv.org/abs/1709.01790}{{\tt 1709.01790}}].

\bibitem{Craig:2019fdy}
N.~Craig, I.~Garcia~Garcia and S.~Koren,  {\em {The Weak Scale from Weak
  Gravity}}, JHEP {\bf 09} (2019) 081
  [\href{http://www.arXiv.org/abs/1904.08426}{{\tt 1904.08426}}].

\bibitem{Draper:2022pvk}
P.~Draper, I.~G. Garcia and M.~Reece,  {\em {Snowmass White Paper: Implications
  of Quantum Gravity for Particle Physics}}, in {\em {2022 Snowmass Summer
  Study}}.
\newblock 3, 2022.
\newblock \href{http://www.arXiv.org/abs/2203.07624}{{\tt 2203.07624}}.

\bibitem{deRham:2022hpx}
C.~de~Rham, S.~Kundu, M.~Reece, A.~J. Tolley and S.-Y. Zhou,  {\em {Snowmass
  White Paper: UV Constraints on IR Physics}}, in {\em {2022 Snowmass Summer
  Study}}.
\newblock 3, 2022.
\newblock \href{http://www.arXiv.org/abs/2203.06805}{{\tt 2203.06805}}.

\bibitem{Cheung:2018cwt}
C.~Cheung, J.~Liu and G.~N. Remmen,  {\em {Proof of the Weak Gravity Conjecture
  from Black Hole Entropy}}, JHEP {\bf 10} (2018) 004
  [\href{http://www.arXiv.org/abs/1801.08546}{{\tt 1801.08546}}].

\bibitem{Hamada:2018dde}
Y.~Hamada, T.~Noumi and G.~Shiu,  {\em {Weak Gravity Conjecture from Unitarity
  and Causality}}, Phys. Rev. Lett. {\bf 123} (2019), no.~5, 051601
  [\href{http://www.arXiv.org/abs/1810.03637}{{\tt 1810.03637}}].

\bibitem{Montero:2018fns}
M.~Montero,  {\em {A Holographic Derivation of the Weak Gravity Conjecture}},
  JHEP {\bf 03} (2019) 157 [\href{http://www.arXiv.org/abs/1812.03978}{{\tt
  1812.03978}}].

\bibitem{Bellazzini:2019xts}
B.~Bellazzini, M.~Lewandowski and J.~Serra,  {\em {Positivity of Amplitudes,
  Weak Gravity Conjecture, and Modified Gravity}}, Phys. Rev. Lett. {\bf 123}
  (2019), no.~25, 251103 [\href{http://www.arXiv.org/abs/1902.03250}{{\tt
  1902.03250}}].

\bibitem{Arkani-Hamed:2021ajd}
N.~Arkani-Hamed, Y.-t. Huang, J.-Y. Liu and G.~N. Remmen,  {\em {Causality,
  unitarity, and the weak gravity conjecture}}, JHEP {\bf 03} (2022) 083
  [\href{http://www.arXiv.org/abs/2109.13937}{{\tt 2109.13937}}].

\bibitem{Polchinski:2003bq}
J.~Polchinski,  {\em {Monopoles, duality, and string theory}}, Int. J. Mod.
  Phys. A {\bf 19S1} (2004) 145--156
  [\href{http://www.arXiv.org/abs/hep-th/0304042}{{\tt hep-th/0304042}}].

\bibitem{Kim:2019vuc}
H.-C. Kim, G.~Shiu and C.~Vafa,  {\em {Branes and the Swampland}}, Phys. Rev. D
  {\bf 100} (2019), no.~6, 066006
  [\href{http://www.arXiv.org/abs/1905.08261}{{\tt 1905.08261}}].

\bibitem{Cvetic:2020kuw}
M.~Cveti\v{c}, M.~Dierigl, L.~Lin and H.~Y. Zhang,  {\em {String Universality
  and Non-Simply-Connected Gauge Groups in 8d}}, Phys. Rev. Lett. {\bf 125}
  (2020), no.~21, 211602 [\href{http://www.arXiv.org/abs/2008.10605}{{\tt
  2008.10605}}].

\bibitem{Montero:2020icj}
M.~Montero and C.~Vafa,  {\em {Cobordism Conjecture, Anomalies, and the String
  Lamppost Principle}}, JHEP {\bf 01} (2021) 063
  [\href{http://www.arXiv.org/abs/2008.11729}{{\tt 2008.11729}}].

\bibitem{Font:2020rsk}
A.~Font, B.~Fraiman, M.~Gra\~na, C.~A. N\'u\~nez and H.~P. De~Freitas,  {\em
  {Exploring the landscape of heterotic strings on $T^d$}}, JHEP {\bf 10}
  (2020) 194 [\href{http://www.arXiv.org/abs/2007.10358}{{\tt 2007.10358}}].

\bibitem{Cvetic:2021sjm}
M.~Cveti\v{c}, M.~Dierigl, L.~Lin and H.~Y. Zhang,  {\em {Gauge group topology
  of 8D Chaudhuri-Hockney-Lykken vacua}}, Phys. Rev. D {\bf 104} (2021), no.~8,
  086018 [\href{http://www.arXiv.org/abs/2107.04031}{{\tt 2107.04031}}].

\bibitem{Font:2021uyw}
A.~Font, B.~Fraiman, M.~Gra\~na, C.~A. N\'u\~nez and H.~Parra De~Freitas,  {\em
  {Exploring the landscape of CHL strings on T$^{d}$}}, JHEP {\bf 08} (2021)
  095 [\href{http://www.arXiv.org/abs/2104.07131}{{\tt 2104.07131}}].

\bibitem{Cvetic:2021sxm}
M.~Cveti\v{c}, M.~Dierigl, L.~Lin and H.~Y. Zhang,  {\em {Higher-form
  symmetries and their anomalies in M-/F-theory duality}}, Phys. Rev. D {\bf
  104} (2021), no.~12, 126019 [\href{http://www.arXiv.org/abs/2106.07654}{{\tt
  2106.07654}}].

\bibitem{Bedroya:2021fbu}
A.~Bedroya, Y.~Hamada, M.~Montero and C.~Vafa,  {\em {Compactness of brane
  moduli and the String Lamppost Principle in d \ensuremath{>} 6}}, JHEP {\bf
  02} (2022) 082 [\href{http://www.arXiv.org/abs/2110.10157}{{\tt
  2110.10157}}].

\bibitem{Cvetic:2022uuu}
M.~Cveti\v{c}, M.~Dierigl, L.~Lin and H.~Y. Zhang,  {\em {One Loop to Rule Them
  All: Eight and Nine Dimensional String Vacua from Junctions}},
  \href{http://www.arXiv.org/abs/2203.03644}{{\tt 2203.03644}}.

\bibitem{Adams:2010zy}
A.~Adams, O.~DeWolfe and W.~Taylor,  {\em {String universality in ten
  dimensions}}, Phys. Rev. Lett. {\bf 105} (2010) 071601
  [\href{http://www.arXiv.org/abs/1006.1352}{{\tt 1006.1352}}].

\bibitem{Morrison:2021wuv}
D.~R. Morrison and W.~Taylor,  {\em {Charge completeness and the massless
  charge lattice in F-theory models of supergravity}}, JHEP {\bf 12} (2021) 040
  [\href{http://www.arXiv.org/abs/2108.02309}{{\tt 2108.02309}}].

\bibitem{Raghuram:2020vxm}
N.~Raghuram, W.~Taylor and A.~P. Turner,  {\em {Automatic enhancement in 6D
  supergravity and F-theory models}}, JHEP {\bf 07} (2021) 048
  [\href{http://www.arXiv.org/abs/2012.01437}{{\tt 2012.01437}}].

\bibitem{Gray:2014fla}
J.~Gray, A.~S. Haupt and A.~Lukas,  {\em {Topological Invariants and Fibration
  Structure of Complete Intersection Calabi-Yau Four-Folds}}, JHEP {\bf 09}
  (2014) 093 [\href{http://www.arXiv.org/abs/1405.2073}{{\tt 1405.2073}}].

\bibitem{Anderson:2017aux}
L.~B. Anderson, X.~Gao, J.~Gray and S.-J. Lee,  {\em {Fibrations in CICY
  Threefolds}}, JHEP {\bf 10} (2017) 077
  [\href{http://www.arXiv.org/abs/1708.07907}{{\tt 1708.07907}}].

\bibitem{Huang:2018esr}
Y.-C. Huang and W.~Taylor,  {\em {On the prevalence of elliptic and genus one
  fibrations among toric hypersurface Calabi-Yau threefolds}}, JHEP {\bf 03}
  (2019) 014 [\href{http://www.arXiv.org/abs/1809.05160}{{\tt 1809.05160}}].

\bibitem{Huang:2019pne}
Y.-C. Huang and W.~Taylor,  {\em {Fibration structure in toric hypersurface
  Calabi-Yau threefolds}}, JHEP {\bf 03} (2020) 172
  [\href{http://www.arXiv.org/abs/1907.09482}{{\tt 1907.09482}}].

\bibitem{DiCerboSvaldi}
G.~{Di Cerbo} and R.~{Svaldi},  {\em {Birational boundedness of low dimensional
  elliptic Calabi-Yau varieties with a section}}, arXiv e-prints (Aug., 2016)
  arXiv:1608.02997 [\href{http://www.arXiv.org/abs/1608.02997}{{\tt
  1608.02997}}].

\bibitem{Grassi}
A.~{Grassi},  {\em {On minimal models of elliptic threefolds}}, Math. Ann. {\bf
  290} (1991) 287--301.

\bibitem{Gross-finite}
M.~{Gross},  {\em {A finiteness theorem for elliptic Calabi-Yau threefolds}},
  Duke Math. Jour. {\bf 74} (1994) 271.

\bibitem{Katz:1996xe}
S.~H. Katz and C.~Vafa,  {\em {Matter from geometry}}, Nucl. Phys. B {\bf 497}
  (1997) 146--154 [\href{http://www.arXiv.org/abs/hep-th/9606086}{{\tt
  hep-th/9606086}}].

\bibitem{Arras:2016evy}
P.~Arras, A.~Grassi and T.~Weigand,  {\em {Terminal Singularities, Milnor
  Numbers, and Matter in F-theory}}, J. Geom. Phys. {\bf 123} (2018) 71--97
  [\href{http://www.arXiv.org/abs/1612.05646}{{\tt 1612.05646}}].

\bibitem{Esole:2017kyr}
M.~Esole, P.~Jefferson and M.~J. Kang,  {\em {Euler Characteristics of Crepant
  Resolutions of Weierstrass Models>}}, Commun. Math. Phys. {\bf 371} (2019),
  no.~1, 99--144 [\href{http://www.arXiv.org/abs/1703.00905}{{\tt
  1703.00905}}].

\bibitem{Acharya:2004qe}
B.~S. Acharya and S.~Gukov,  {\em {M theory and singularities of exceptional
  holonomy manifolds}}, Phys. Rept. {\bf 392} (2004) 121--189
  [\href{http://www.arXiv.org/abs/hep-th/0409191}{{\tt hep-th/0409191}}].

\bibitem{Kovalev:2001zr}
A.~Kovalev,  {\em {Twisted connected sums and special Riemannian holonomy}},
  \href{http://www.arXiv.org/abs/math/0012189}{{\tt math/0012189}}.

\bibitem{Corti:2012kd}
A.~Corti, M.~Haskins, J.~Nordstr\"om and T.~Pacini,  {\em
  {$\mathrm{G}_{2}$-manifolds and associative submanifolds via semi-Fano
  $3$-folds}}, Duke Math. J. {\bf 164} (2015), no.~10, 1971--2092
  [\href{http://www.arXiv.org/abs/1207.4470}{{\tt 1207.4470}}].

\bibitem{Halverson:2014tya}
J.~Halverson and D.~R. Morrison,  {\em {The landscape of M-theory
  compactifications on seven-manifolds with G$_{2}$ holonomy}}, JHEP {\bf 04}
  (2015) 047 [\href{http://www.arXiv.org/abs/1412.4123}{{\tt 1412.4123}}].

\bibitem{Halverson:2015vta}
J.~Halverson and D.~R. Morrison,  {\em {On gauge enhancement and singular
  limits in G$_{2}$ compactifications of M-theory}}, JHEP {\bf 04} (2016) 100
  [\href{http://www.arXiv.org/abs/1507.05965}{{\tt 1507.05965}}].

\bibitem{Braun:2016igl}
A.~P. Braun,  {\em {Tops as building blocks for G$_{2}$ manifolds}}, JHEP {\bf
  10} (2017) 083 [\href{http://www.arXiv.org/abs/1602.03521}{{\tt
  1602.03521}}].

\bibitem{daCGuio:2017ifs}
T.~C. da~C.~Guio, H.~Jockers, A.~Klemm and H.-Y. Yeh,  {\em {Effective Action
  from M-Theory on Twisted Connected Sum G$_{2}$-Manifolds}}, Commun. Math.
  Phys. {\bf 359} (2018), no.~2, 535--601
  [\href{http://www.arXiv.org/abs/1702.05435}{{\tt 1702.05435}}].

\bibitem{Braun:2017ryx}
A.~P. Braun and M.~Del~Zotto,  {\em {Mirror Symmetry for $G_2$-Manifolds:
  Twisted Connected Sums and Dual Tops}}, JHEP {\bf 05} (2017) 080
  [\href{http://www.arXiv.org/abs/1701.05202}{{\tt 1701.05202}}].

\bibitem{Braun:2017csz}
A.~P. Braun and M.~Del~Zotto,  {\em {Towards Generalized Mirror Symmetry for
  Twisted Connected Sum $G_2$ Manifolds}}, JHEP {\bf 03} (2018) 082
  [\href{http://www.arXiv.org/abs/1712.06571}{{\tt 1712.06571}}].

\bibitem{Braun:2018fdp}
A.~P. Braun, M.~Del~Zotto, J.~Halverson, M.~Larfors, D.~R. Morrison and
  S.~Sch\"afer-Nameki,  {\em {Infinitely many M2-instanton corrections to
  M-theory on G$_{2}$-manifolds}}, JHEP {\bf 09} (2018) 077
  [\href{http://www.arXiv.org/abs/1803.02343}{{\tt 1803.02343}}].

\bibitem{Pantev:2009de}
T.~Pantev and M.~Wijnholt,  {\em {Hitchin's Equations and M-Theory
  Phenomenology}}, J. Geom. Phys. {\bf 61} (2011) 1223--1247
  [\href{http://www.arXiv.org/abs/0905.1968}{{\tt 0905.1968}}].

\bibitem{Braun:2018vhk}
A.~P. Braun, S.~Cizel, M.~H\"ubner and S.~Sch\"afer-Nameki,  {\em {Higgs
  bundles for M-theory on $G_{2}$-manifolds}}, JHEP {\bf 03} (2019) 199
  [\href{http://www.arXiv.org/abs/1812.06072}{{\tt 1812.06072}}].

\bibitem{Barbosa:2019bgh}
R.~Barbosa, M.~Cveti\v{c}, J.~J. Heckman, C.~Lawrie, E.~Torres and
  G.~Zoccarato,  {\em {T-branes and $G_2$ backgrounds}}, Phys. Rev. D {\bf 101}
  (2020), no.~2, 026015 [\href{http://www.arXiv.org/abs/1906.02212}{{\tt
  1906.02212}}].

\bibitem{Hubner:2020yde}
M.~Hubner,  {\em {Local G$_{2}$-manifolds, Higgs bundles and a colored quantum
  mechanics}}, JHEP {\bf 05} (2021) 002
  [\href{http://www.arXiv.org/abs/2009.07136}{{\tt 2009.07136}}].

\bibitem{Cvetic:2020piw}
M.~Cveti\v{c}, J.~J. Heckman, T.~B. Rochais, E.~Torres and G.~Zoccarato,  {\em
  {Geometric unification of Higgs bundle vacua}}, Phys. Rev. D {\bf 102}
  (2020), no.~10, 106012 [\href{http://www.arXiv.org/abs/2003.13682}{{\tt
  2003.13682}}].

\bibitem{Acharya:2020vmg}
B.~S. Acharya, L.~Foscolo, M.~Najjar and E.~E. Svanes,  {\em {New
  G$_{2}$-conifolds in M-theory and their field theory interpretation}}, JHEP
  {\bf 05} (2021) 250 [\href{http://www.arXiv.org/abs/2011.06998}{{\tt
  2011.06998}}].

\bibitem{Cvetic:2022imb}
M.~Cveti\v{c}, J.~J. Heckman, M.~H\"ubner and E.~Torres,  {\em {0-Form, 1-Form
  and 2-Group Symmetries via Cutting and Gluing of Orbifolds}},
  \href{http://www.arXiv.org/abs/2203.10102}{{\tt 2203.10102}}.

\bibitem{Grassi:2018wfy}
A.~Grassi, J.~Halverson, C.~Long, J.~L. Shaneson and J.~Tian,  {\em
  {Non-simply-laced Symmetry Algebras in F-theory on Singular Spaces}}, JHEP
  {\bf 09} (2018) 129 [\href{http://www.arXiv.org/abs/1805.06949}{{\tt
  1805.06949}}].

\bibitem{Grassi:2021ptc}
A.~Grassi, J.~Halverson, C.~Long, J.~L. Shaneson, B.~Sung and J.~Tian,  {\em
  {$6$D Anomaly-Free Matter Spectrum in F-theory on Singular Spaces}},
  \href{http://www.arXiv.org/abs/2110.06943}{{\tt 2110.06943}}.

\bibitem{Cvetic:2001kk}
M.~Cveti{\v c}, G.~Shiu and A.~M. Uranga,  {\em {Chiral type II orientifold
  constructions as M theory on G(2) holonomy spaces}}, in {\em {9th
  International Conference on Supersymmetry and Unification of Fundamental
  Interactions (SUSY01)}}, pp.~317--326.
\newblock 11, 2001.
\newblock \href{http://www.arXiv.org/abs/hep-th/0111179}{{\tt hep-th/0111179}}.

\bibitem{Acharya:2001gy}
B.~S. Acharya and E.~Witten,  {\em {Chiral fermions from manifolds of G(2)
  holonomy}}, \href{http://www.arXiv.org/abs/hep-th/0109152}{{\tt
  hep-th/0109152}}.

\bibitem{GORESKY1980135}
M.~Goresky and R.~MacPherson,  {\em Intersection homology theory}, Topology
  {\bf 19} (1980), no.~2, 135--162.

\bibitem{Headrick:2005ch}
M.~Headrick and T.~Wiseman,  {\em {Numerical Ricci-flat metrics on K3}}, Class.
  Quant. Grav. {\bf 22} (2005) 4931--4960
  [\href{http://www.arXiv.org/abs/hep-th/0506129}{{\tt hep-th/0506129}}].

\bibitem{Donaldsonnumerical}
S.~K. Donaldson,  {\em {Some numerical results in complex differential
  geometry}}, \href{http://www.arXiv.org/abs/math/0512625}{{\tt math/0512625}}.

\bibitem{Douglas:2006rr}
M.~R. Douglas, R.~L. Karp, S.~Lukic and R.~Reinbacher,  {\em {Numerical
  Calabi-Yau metrics}}, J. Math. Phys. {\bf 49} (2008) 032302
  [\href{http://www.arXiv.org/abs/hep-th/0612075}{{\tt hep-th/0612075}}].

\bibitem{Braun:2007sn}
V.~Braun, T.~Brelidze, M.~R. Douglas and B.~A. Ovrut,  {\em {Calabi-Yau Metrics
  for Quotients and Complete Intersections}}, JHEP {\bf 05} (2008) 080
  [\href{http://www.arXiv.org/abs/0712.3563}{{\tt 0712.3563}}].

\bibitem{Anderson:2010ke}
L.~B. Anderson, V.~Braun, R.~L. Karp and B.~A. Ovrut,  {\em {Numerical
  Hermitian Yang-Mills Connections and Vector Bundle Stability in Heterotic
  Theories}}, JHEP {\bf 06} (2010) 107
  [\href{http://www.arXiv.org/abs/1004.4399}{{\tt 1004.4399}}].

\bibitem{Kachru:2020tat}
S.~Kachru, A.~Tripathy and M.~Zimet,  {\em {K3 metrics}},
  \href{http://www.arXiv.org/abs/2006.02435}{{\tt 2006.02435}}.

\bibitem{Lerche:1986cx}
W.~Lerche, D.~Lust and A.~N. Schellekens,  {\em {Chiral Four-Dimensional
  Heterotic Strings from Selfdual Lattices}}, Nucl. Phys. B {\bf 287} (1987)
  477.

\bibitem{Bousso:2000xa}
R.~Bousso and J.~Polchinski,  {\em {Quantization of four form fluxes and
  dynamical neutralization of the cosmological constant}}, JHEP {\bf 06} (2000)
  006 [\href{http://www.arXiv.org/abs/hep-th/0004134}{{\tt hep-th/0004134}}].

\bibitem{Weinberg:1987dv}
S.~Weinberg,  {\em {Anthropic Bound on the Cosmological Constant}}, Phys. Rev.
  Lett. {\bf 59} (1987) 2607.

\bibitem{Scholler:2018apc}
F.~Sch\"oller and H.~Skarke,  {\em {All Weight Systems for
  Calabi\textendash{}Yau Fourfolds from Reflexive Polyhedra}}, Commun. Math.
  Phys. {\bf 372} (2019), no.~2, 657--678
  [\href{http://www.arXiv.org/abs/1808.02422}{{\tt 1808.02422}}].

\bibitem{Morrison:2012js}
D.~R. Morrison and W.~Taylor,  {\em {Toric bases for 6D F-theory models}},
  Fortsch. Phys. {\bf 60} (2012) 1187--1216
  [\href{http://www.arXiv.org/abs/1204.0283}{{\tt 1204.0283}}].

\bibitem{Taylor:2015isa}
W.~Taylor and Y.-N. Wang,  {\em {Non-toric bases for elliptic
  Calabi\textendash{}Yau threefolds and 6D F-theory vacua}}, Adv. Theor. Math.
  Phys. {\bf 21} (2017) 1063--1114
  [\href{http://www.arXiv.org/abs/1504.07689}{{\tt 1504.07689}}].

\bibitem{Anderson:2014gla}
L.~B. Anderson and W.~Taylor,  {\em {Geometric constraints in dual F-theory and
  heterotic string compactifications}}, JHEP {\bf 08} (2014) 025
  [\href{http://www.arXiv.org/abs/1405.2074}{{\tt 1405.2074}}].

\bibitem{Halverson:2015jua}
J.~Halverson and W.~Taylor,  {\em {$ {\mathrm{\mathbb{P}}}^1 $-bundle bases and
  the prevalence of non-Higgsable structure in 4D F-theory models}}, JHEP {\bf
  09} (2015) 086 [\href{http://www.arXiv.org/abs/1506.03204}{{\tt
  1506.03204}}].

\bibitem{Denef:2006ad}
F.~Denef and M.~R. Douglas,  {\em {Computational complexity of the landscape.
  I.}}, Annals Phys. {\bf 322} (2007) 1096--1142
  [\href{http://www.arXiv.org/abs/hep-th/0602072}{{\tt hep-th/0602072}}].

\bibitem{Halverson:2018cio}
J.~Halverson and F.~Ruehle,  {\em {Computational Complexity of Vacua and
  Near-Vacua in Field and String Theory}}, Phys. Rev. D {\bf 99} (2019), no.~4,
  046015 [\href{http://www.arXiv.org/abs/1809.08279}{{\tt 1809.08279}}].

\bibitem{Cvetic:2010ky}
M.~Cveti\v{c}, I.~Garcia-Etxebarria and J.~Halverson,  {\em {On the computation
  of non-perturbative effective potentials in the string theory landscape:
  IIB/F-theory perspective}}, Fortsch. Phys. {\bf 59} (2011) 243--283
  [\href{http://www.arXiv.org/abs/1009.5386}{{\tt 1009.5386}}].

\bibitem{Halverson:2019vmd}
J.~Halverson, M.~Plesser, F.~Ruehle and J.~Tian,  {\em {K\"ahler Moduli
  Stabilization and the Propagation of Decidability}}, Phys. Rev. D {\bf 101}
  (2020), no.~4, 046010 [\href{http://www.arXiv.org/abs/1911.07835}{{\tt
  1911.07835}}].

\bibitem{Bao:2017thx}
N.~Bao, R.~Bousso, S.~Jordan and B.~Lackey,  {\em {Fast optimization algorithms
  and the cosmological constant}}, Phys. Rev. D {\bf 96} (2017), no.~10, 103512
  [\href{http://www.arXiv.org/abs/1706.08503}{{\tt 1706.08503}}].

\bibitem{He:2017aed}
Y.-H. He,  {\em {Deep-Learning the Landscape}},
  \href{http://www.arXiv.org/abs/1706.02714}{{\tt 1706.02714}}.

\bibitem{Ruehle:2017mzq}
F.~Ruehle,  {\em {Evolving neural networks with genetic algorithms to study the
  String Landscape}}, JHEP {\bf 08} (2017) 038
  [\href{http://www.arXiv.org/abs/1706.07024}{{\tt 1706.07024}}].

\bibitem{Krefl:2017yox}
D.~Krefl and R.-K. Seong,  {\em {Machine Learning of Calabi-Yau Volumes}},
  Phys. Rev. D {\bf 96} (2017), no.~6, 066014
  [\href{http://www.arXiv.org/abs/1706.03346}{{\tt 1706.03346}}].

\bibitem{Carifio:2017bov}
J.~Carifio, J.~Halverson, D.~Krioukov and B.~D. Nelson,  {\em {Machine Learning
  in the String Landscape}}, JHEP {\bf 09} (2017) 157
  [\href{http://www.arXiv.org/abs/1707.00655}{{\tt 1707.00655}}].

\bibitem{cirafici2016persistent}
M.~Cirafici,  {\em Persistent homology and string vacua}, Journal of High
  Energy Physics {\bf 2016} (2016), no.~3, 1--34.

\bibitem{Cole:2018emh}
A.~Cole and G.~Shiu,  {\em {Topological Data Analysis for the String
  Landscape}}, JHEP {\bf 03} (2019) 054
  [\href{http://www.arXiv.org/abs/1812.06960}{{\tt 1812.06960}}].

\bibitem{Carifio:2017nyb}
J.~Carifio, W.~J. Cunningham, J.~Halverson, D.~Krioukov, C.~Long and B.~D.
  Nelson,  {\em {Vacuum Selection from Cosmology on Networks of String
  Geometries}}, Phys. Rev. Lett. {\bf 121} (2018), no.~10, 101602
  [\href{http://www.arXiv.org/abs/1711.06685}{{\tt 1711.06685}}].

\bibitem{Faraggi:2021mws}
A.~E. Faraggi, B.~Percival, S.~Schewe and D.~Wojtczak,  {\em {Satisfiability
  modulo theories and chiral heterotic string vacua with positive cosmological
  constant}}, Phys. Lett. B {\bf 816} (2021) 136187
  [\href{http://www.arXiv.org/abs/2101.03227}{{\tt 2101.03227}}].

\bibitem{Faraggi:2022hut}
A.~E. Faraggi, V.~G. Matyas and B.~Percival,  {\em {Towards Classification of
  $\mathcal{N}=1$ and $\mathcal{N}=0$ Flipped $SU(5)$ Asymmetric $\mathbb{Z}_2
  \times \mathbb{Z}_2$ Heterotic String Orbifolds}},
  \href{http://www.arXiv.org/abs/2202.04507}{{\tt 2202.04507}}.

\bibitem{Blaback:2013ht}
J.~Bl\r{a}b\"ack, U.~Danielsson and G.~Dibitetto,  {\em {Fully stable dS vacua
  from generalised fluxes}}, JHEP {\bf 08} (2013) 054
  [\href{http://www.arXiv.org/abs/1301.7073}{{\tt 1301.7073}}].

\bibitem{Blaback:2013fca}
J.~Bl\r{a}b\"ack, U.~Danielsson and G.~Dibitetto,  {\em {Accelerated Universes
  from type IIA Compactifications}}, JCAP {\bf 03} (2014) 003
  [\href{http://www.arXiv.org/abs/1310.8300}{{\tt 1310.8300}}].

\bibitem{Abel:2014xta}
S.~Abel and J.~Rizos,  {\em {Genetic Algorithms and the Search for Viable
  String Vacua}}, JHEP {\bf 08} (2014) 010
  [\href{http://www.arXiv.org/abs/1404.7359}{{\tt 1404.7359}}].

\bibitem{AbdusSalam:2020ywo}
S.~AbdusSalam, S.~Abel, M.~Cicoli, F.~Quevedo and P.~Shukla,  {\em {A
  systematic approach to K\"ahler moduli stabilisation}}, JHEP {\bf 08} (2020),
  no.~08, 047 [\href{http://www.arXiv.org/abs/2005.11329}{{\tt 2005.11329}}].

\bibitem{Loges:2021hvn}
G.~J. Loges and G.~Shiu,  {\em {Breeding realistic D-brane models}},
  \href{http://www.arXiv.org/abs/2112.08391}{{\tt 2112.08391}}.

\bibitem{Liu:2017dzi}
J.~Liu,  {\em {Artificial Neural Network in Cosmic Landscape}}, JHEP {\bf 12}
  (2017) 149 [\href{http://www.arXiv.org/abs/1707.02800}{{\tt 1707.02800}}].

\bibitem{Wang:2018rkk}
Y.-N. Wang and Z.~Zhang,  {\em {Learning non-Higgsable gauge groups in 4D
  F-theory}}, JHEP {\bf 08} (2018) 009
  [\href{http://www.arXiv.org/abs/1804.07296}{{\tt 1804.07296}}].

\bibitem{Altman:2018zlc}
R.~Altman, J.~Carifio, J.~Halverson and B.~D. Nelson,  {\em {Estimating
  Calabi-Yau Hypersurface and Triangulation Counts with Equation Learners}},
  JHEP {\bf 03} (2019) 186 [\href{http://www.arXiv.org/abs/1811.06490}{{\tt
  1811.06490}}].

\bibitem{Jinno:2018dek}
R.~Jinno,  {\em {Machine learning for bounce calculation}},
  \href{http://www.arXiv.org/abs/1805.12153}{{\tt 1805.12153}}.

\bibitem{Bull:2018uow}
K.~Bull, Y.-H. He, V.~Jejjala and C.~Mishra,  {\em {Machine Learning CICY
  Threefolds}}, Phys. Lett. B {\bf 785} (2018) 65--72
  [\href{http://www.arXiv.org/abs/1806.03121}{{\tt 1806.03121}}].

\bibitem{Rudelius:2018yqi}
T.~Rudelius,  {\em {Learning to Inflate}}, JCAP {\bf 02} (2019) 044
  [\href{http://www.arXiv.org/abs/1810.05159}{{\tt 1810.05159}}].

\bibitem{Jejjala:2019kio}
V.~Jejjala, A.~Kar and O.~Parrikar,  {\em {Deep Learning the Hyperbolic Volume
  of a Knot}}, Phys. Lett. B {\bf 799} (2019) 135033
  [\href{http://www.arXiv.org/abs/1902.05547}{{\tt 1902.05547}}].

\bibitem{Brodie:2019dfx}
C.~R. Brodie, A.~Constantin, R.~Deen and A.~Lukas,  {\em {Machine Learning Line
  Bundle Cohomology}}, Fortsch. Phys. {\bf 68} (2020), no.~1, 1900087
  [\href{http://www.arXiv.org/abs/1906.08730}{{\tt 1906.08730}}].

\bibitem{Bies:2020gvf}
M.~Bies, M.~Cveti\v{c}, R.~Donagi, L.~Lin, M.~Liu and F.~Ruehle,  {\em {Machine
  Learning and Algebraic Approaches towards Complete Matter Spectra in 4d
  F-theory}}, JHEP {\bf 01} (2021) 196
  [\href{http://www.arXiv.org/abs/2007.00009}{{\tt 2007.00009}}].

\bibitem{Halverson:2019tkf}
J.~Halverson, B.~Nelson and F.~Ruehle,  {\em {Branes with Brains: Exploring
  String Vacua with Deep Reinforcement Learning}}, JHEP {\bf 06} (2019) 003
  [\href{http://www.arXiv.org/abs/1903.11616}{{\tt 1903.11616}}].

\bibitem{Larfors:2020ugo}
M.~Larfors and R.~Schneider,  {\em {Explore and Exploit with Heterotic Line
  Bundle Models}}, Fortsch. Phys. {\bf 68} (2020), no.~5, 2000034
  [\href{http://www.arXiv.org/abs/2003.04817}{{\tt 2003.04817}}].

\bibitem{Krippendorf:2021uxu}
S.~Krippendorf, R.~Kroepsch and M.~Syvaeri,  {\em {Revealing systematics in
  phenomenologically viable flux vacua with reinforcement learning}},
  \href{http://www.arXiv.org/abs/2107.04039}{{\tt 2107.04039}}.

\bibitem{Constantin:2021for}
A.~Constantin, T.~R. Harvey and A.~Lukas,  {\em {Heterotic String Model
  Building with Monad Bundles and Reinforcement Learning}},
  \href{http://www.arXiv.org/abs/2108.07316}{{\tt 2108.07316}}.

\bibitem{Silver2017}
D.~Silver, J.~Schrittwieser, K.~Simonyan, I.~Antonoglou, A.~Huang, A.~Guez,
  T.~Hubert, L.~Baker, M.~Lai, A.~Bolton, Y.~Chen, T.~Lillicrap, F.~Hui,
  L.~Sifre, G.~van~den Driessche, T.~Graepel and D.~Hassabis,  {\em Mastering
  the game of Go without human knowledge}, Nature {\bf 550} (Oct, 2017)
  354--359.

\bibitem{Cole:2021nnt}
A.~Cole, S.~Krippendorf, A.~Schachner and G.~Shiu,  {\em {Probing the Structure
  of String Theory Vacua with Genetic Algorithms and Reinforcement Learning}},
  in {\em {35th Conference on Neural Information Processing Systems}}.
\newblock 11, 2021.
\newblock \href{http://www.arXiv.org/abs/2111.11466}{{\tt 2111.11466}}.

\bibitem{Abel:2021ddu}
S.~Abel, A.~Constantin, T.~R. Harvey and A.~Lukas,  {\em {String Model
  Building, Reinforcement Learning and Genetic Algorithms}}, in {\em {Nankai
  Symposium on Mathematical Dialogues}: {In celebration of S.S.Chern's 110th
  anniversary}}.
\newblock 11, 2021.
\newblock \href{http://www.arXiv.org/abs/2111.07333}{{\tt 2111.07333}}.

\bibitem{Abel:2021rrj}
S.~Abel, A.~Constantin, T.~R. Harvey and A.~Lukas,  {\em {Evolving Heterotic
  Gauge Backgrounds: Genetic Algorithms versus Reinforcement Learning}},
  \href{http://www.arXiv.org/abs/2110.14029}{{\tt 2110.14029}}.

\bibitem{Erbin:2018csv}
H.~Erbin and S.~Krippendorf,  {\em {GANs for generating EFT models}}, Phys.
  Lett. B {\bf 810} (2020) 135798
  [\href{http://www.arXiv.org/abs/1809.02612}{{\tt 1809.02612}}].

\bibitem{Halverson:2020opj}
J.~Halverson and C.~Long,  {\em {Statistical Predictions in String Theory and
  Deep Generative Models}}, Fortsch. Phys. {\bf 68} (2020), no.~5, 2000005
  [\href{http://www.arXiv.org/abs/2001.00555}{{\tt 2001.00555}}].

\bibitem{Anderson:2020hux}
L.~B. Anderson, M.~Gerdes, J.~Gray, S.~Krippendorf, N.~Raghuram and F.~Ruehle,
  {\em {Moduli-dependent Calabi-Yau and SU(3)-structure metrics from Machine
  Learning}}, JHEP {\bf 05} (2021) 013
  [\href{http://www.arXiv.org/abs/2012.04656}{{\tt 2012.04656}}].

\bibitem{Douglas:2020hpv}
M.~R. Douglas, S.~Lakshminarasimhan and Y.~Qi,  {\em {Numerical Calabi-Yau
  metrics from holomorphic networks}},
  \href{http://www.arXiv.org/abs/2012.04797}{{\tt 2012.04797}}.

\bibitem{Jejjala:2020wcc}
V.~Jejjala, D.~K. Mayorga~Pena and C.~Mishra,  {\em {Neural Network
  Approximations for Calabi-Yau Metrics}},
  \href{http://www.arXiv.org/abs/2012.15821}{{\tt 2012.15821}}.

\bibitem{Larfors:2021pbb}
M.~Larfors, A.~Lukas, F.~Ruehle and R.~Schneider,  {\em {Learning Size and
  Shape of Calabi-Yau Spaces}}, \href{http://www.arXiv.org/abs/2111.01436}{{\tt
  2111.01436}}.

\bibitem{Ashmore:2019wzb}
A.~Ashmore, Y.-H. He and B.~A. Ovrut,  {\em {Machine Learning
  Calabi\textendash{}Yau Metrics}}, Fortsch. Phys. {\bf 68} (2020), no.~9,
  2000068 [\href{http://www.arXiv.org/abs/1910.08605}{{\tt 1910.08605}}].

\bibitem{2017Sci...355..602C}
G.~{Carleo} and M.~{Troyer},  {\em {Solving the quantum many-body problem with
  artificial neural networks}}, Science {\bf 355} (Feb., 2017) 602--606
  [\href{http://www.arXiv.org/abs/1606.02318}{{\tt 1606.02318}}].

\bibitem{RUEHLE20201}
F.~Ruehle,  {\em Data science applications to string theory}, Physics Reports
  {\bf 839} (2020) 1--117 Data science applications to string theory.

\end{thebibliography}\endgroup

\end{document}